\documentclass[12pt,oneside,bibliography=totoc,abstracton]{scrartcl}

\usepackage{geometry}
\usepackage[english]{babel} % Document in English language
\usepackage{graphicx}% More options when including graphics
\usepackage[utf8]{inputenc}% Correct font encoding for special characters
\usepackage[T1]{fontenc}% Correct separation for special characters
\usepackage{lmodern}% Modern font type

\usepackage{array}% More table options
\usepackage[table]{xcolor}% Table color options
\usepackage{ragged2e}% More table options
\usepackage{verbatim}% New environment, also for comments

\usepackage{fancyhdr}% Can create headers and footers
\usepackage{amssymb}% Provides usage of special characters and symbols
\usepackage{amsthm}% Offers theorem environment
\usepackage{amsmath}% Provides mathematical symbols
\usepackage{stmaryrd}% More mathematical symbols
\usepackage{bm}% Bold in math mode
\usepackage[ruled,vlined,linesnumbered,resetcount]{algorithm2e}% Algorithms
\usepackage{color}% Colored text

\usepackage{subcaption}% Makes it possible to create sub captions in tables
\usepackage{setspace}% Makes it possible to change spacing
\usepackage{textcomp}% Font that provides more special characters
\usepackage{lscape}% Modifies margins
\usepackage{wasysym}% More symbols

\usepackage{hyperref}% Appearance of links do not get changed

\definecolor{purple}{RGB}{206,0,206}

\hypersetup{
	colorlinks,
	citecolor=purple,
	linkcolor=red,
	urlcolor=blue
}

\newcommand{\zz}{$\mathrm{T\kern-.4em\raise-0.5ex\hbox{P}}$}
\newcommand{\pto}{\mathrel{\ooalign{\hfil$\mapstochar$\hfil\cr$\to$\cr}}}
\newcommand{\ito}{\xrightarrow{\,\smash{\raisebox{-0.45ex}{\ensuremath{\scriptstyle\sim}}}\,}}
\DeclareMathOperator{\lift}{lift}
\newcommand{\emptyccell}{\multicolumn{1}{c}{--}}
\newcommand{\emptyccellrightended}{\multicolumn{1}{c|}{--}}
\newcommand{\figref}[1]{\textbf{Fig.~\ref{#1}}}
\newcommand{\tableref}[1]{\textbf{Table~\ref{#1}}}
\newcommand{\lemmaref}[1]{\textbf{Lemma~\ref{#1}}}
\newcommand{\theoremref}[1]{\textbf{Theorem~\ref{#1}}}
\newcommand{\corollaryref}[1]{\textbf{Corollary~\ref{#1}}}

\newcommand{\propositionref}[1]{\textbf{Proposition~\ref{#1}}}
\newcommand{\sectionref}[1]{\textbf{Section~\ref{#1}}}
\newcommand{\algoref}[1]{\textbf{Algorithm~\ref{#1}}}
\newcommand{\defref}[1]{\textbf{Definition~\ref{#1}}}
\newcommand{\libref}[1]{\textbf{\cite{#1}}}
\newtheorem{mydef}{Definition}
\newtheorem{mylemma}{Lemma}
\newtheorem{mytheorem}{Theorem}
\newtheorem{mycorollary}{Colorally}
\newtheorem{myproposition}{Proposition}

\begin{document}
% Command renews that need to be inside the document
\renewcommand{\figurename}{Fig.}% Fig. instead of Figure. when linking figures
% Titlepage
\begin{titlepage}
	\title{Minimization of Büchi Automata using Fair Simulation}
	\subtitle{Bachelorthesis} 
	\author{Daniel Tischner}
	\publishers{
		\begin{tabular}{ll}
			\\
			Supervisor:	& Prof.~Dr.~Andreas Podelski \\
			Advisor:		&  Matthias Heizmann 
		\end{tabular}
	}
	\maketitle
	% Suppress page number on title page
	\thispagestyle{empty}
\end{titlepage}
\newpage
{
	\hypersetup{linkcolor=black}
	\tableofcontents
}
\clearpage
% Declaration
\renewcommand{\abstractname}{\Huge Declaration}
\begin{abstract}
	\vbox{}
	I hereby declare, that I am the sole author and composer of my Thesis and that
	no other sources or learning aids, other than those listed, have been used.
	Furthermore, I declare that I have acknowledged the work of others by providing
	detailed references of said work. I hereby also declare, that my Thesis has not
	been prepared for another examination or assignment, either
	wholly or excerpts thereof.
	\vfill
	\parbox{4cm}{\hrule
	\strut \centering\footnotesize Place, Date} \hfill\parbox{4cm}{\hrule
	\strut \centering\footnotesize Signature}
\end{abstract}
\clearpage
% Zusammenfassung
\renewcommand{\abstractname}{\Huge Zusammenfassung}
\begin{abstract}\quad\\
	Wir präsentieren einen Algorithmus, der die Größe von Büchi Automaten
	mit Hilfe von \textit{fair simulation} reduziert.
	Seine Zeitkomplexität ist $\mathcal{O}(|Q|^4 \cdot |\Delta|^2)$, die Platzkomplexität ist
	$\mathcal{O}(|Q| \cdot |\Delta|)$.
	
	Simulation ist ein häufig genutzter Ansatz zur Minimierung von $\omega$-Automaten,
	wie Büchi Automaten. \textit{Direct simulation}, \textit{delayed simulation}
	und \textit{fair simulation} sind verschiedene Simulationstypen.
	Wie wir zeigen werden, ist Minimierung basierend auf \textit{direct} oder
	\textit{delayed simulation} konzeptionell einfach. Wohingegen der Algorithmus
	basierend auf \textit{fair simulation} komplexer ist.
	Allerdings erlaubt \textit{fair simulation} eine stärkere Minimierung des Automaten.
	
	Des Weiteren erläutern wir die Theorie hinter dem Algorithmus, umfassen in der Praxis nützliche
	Optimierungen, zeigen Versuchsergebnisse auf und vergleichen unsere
	Technik mit anderen Minimierungsstrategien.
\end{abstract}
\clearpage
% Abstract
\renewcommand{\abstractname}{\Huge Abstract}
\begin{abstract}\quad\\
	We present an algorithm, which reduces the size of Büchi automata
	using \textit{fair simulation}.
	Its time complexity is $\mathcal{O}(|Q|^4 \cdot |\Delta|^2)$,
	the space complexity is $\mathcal{O}(|Q| \cdot |\Delta|)$.
	
	Simulation is a common approach for minimizing $\omega$-automata
	such as Büchi automata. \textit{Direct simulation}, \textit{delayed simulation}
	and \textit{fair simulation} are different types of simulation.
	As we will show, minimization based on direct or delayed simulation is conceptually
	simple. Whereas the algorithm based on fair simulation is more complex.
	However, fair simulation allows a stronger minimization of the automaton.
	
	Further, we illustrate the theory behind the algorithm, cover optimizations
	useful in practice, give experimental results and compare our technique to
	other minimization strategies.
\end{abstract}
\clearpage
% Commands that need to be after the preamble
\fancyhead[LO,RE]{\parbox{0.7\textwidth}{\small{\textbf{\rightmark}}}}
\fancyhead[LE,RO]{\textbf{\small{Section \arabic{section}}}}
% Introduction
\section{Introduction}
Many applications heavily make use of automata, as shown in
\libref{fair_minimization, advanced_minimization, ltl_formulae}.
Büchi automata, alongside LTL, are commonly used for model checking.
Efficiently reducing the size of automata without changing their language
greatly improves the capabilities of such applications, since the
amount of states and transitions often form a bottleneck.\\
% Figures
\begin{figure}[ht]
	 \begin{center}
		\includegraphics[scale=0.5]{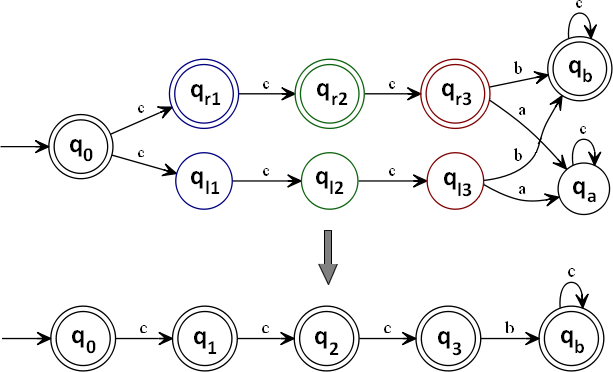}
	\end{center}
	\caption{Automaton where there is no mutually delayed simulation
		but three pairs of states that \textit{fairly simulate} each other,
		$\{(q_{ri}, q_{li}) | i \in \mathbb{N}\}$. The algorithm presented in this thesis
		(see \sectionref{section_minimization}) checks whether merging those states changes
		the language. Since it does not, the merges are accepted.
		The state $q_a$ gets removed because $cccac^{\omega}$ is
		no \textit{accepting word}.}
	\label{fair_nodelayed1}
\end{figure}
\begin{figure}
	\begin{center}
		\includegraphics[scale=0.5]{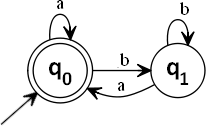}
	\end{center}
	\caption{This automaton, which accepts the language
		$\mathcal{L}(A) = \{w \in \Sigma^{\omega} | w = (b^*a)^\omega\}$,
		cannot be minimized without changing the language, although the two
		states fairly simulate each other.}
	\label{fair_fail}\quad\\
\end{figure}
\begin{figure}
	 \begin{center}
		\includegraphics[scale=0.8]{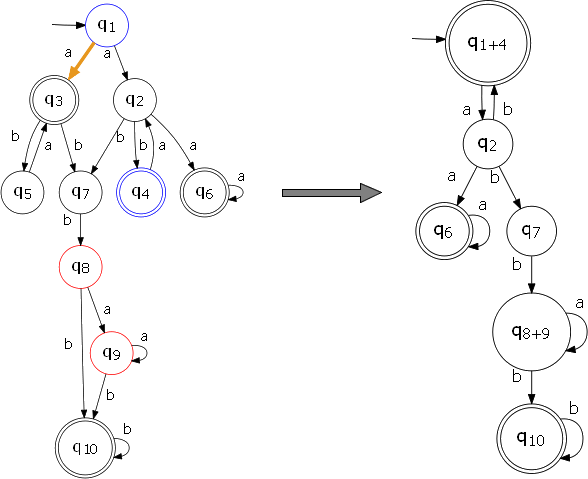}
	\end{center}
	\caption{An example automaton, taken from \libref{fair_minimization}, where there is no mutually delayed simulation,
		but two safely mergeable, fairly simulating pairs of states, $(q_1, q_4)$ and $(q_8, q_9)$.
		Further, the algorithm presented in this thesis (see \sectionref{section_minimization})
		removes the transition $(q_1, a, q_3)$, which turns out to be redundant for the language of the automaton.
		States $q_3$ and $q_5$ become unreachable and are also removed.}
	\label{fair_nodelayed2}\quad\\
\end{figure}\quad\\
The term \textit{minimizing an automaton} can be used in different ways. For $\omega$-automata, finding a
language-equivalent automaton with minimal size is \textit{NP-hard}. This was proven in \libref{nphard}.
Instead, in the context of this thesis, it is defined as the search for any language-equivalent
automaton with less states and transitions.\\\\
A common approach for minimization consists of finding pairs of states that can be merged without
changing the language of the automaton. They are called merge-equivalent. But how can those efficiently
be found for $\omega$-automata such as Büchi automata?

Many existing solutions are based on \textit{simulation}. Moreover, as seen
in \libref{simulation_general}, they use \textit{direct simulation} or \textit{delayed simulation}.
Simulation provides pairs of states, they are called simulation-equivalent.
For direct and delayed simulation such pairs are always also merge-equivalent, as shown in \libref{simulation_general}.
Unfortunately, these types of simulation are not always practicable. For delayed simulation,
they quickly run out of space for big automata. For direct simulation, they just remove few states.\\\\
In this work, we present an algorithm first introduced in \libref{fair_minimization}
which reduces the size of Büchi automata using \textit{fair simulation}.
Its time complexity is $\mathcal{O}(|Q|^4 \cdot |\Delta|^2)$, the space complexity is $\mathcal{O}(|Q| \cdot |\Delta|)$.
The state pairs provided by fair simulation are a superset of the pairs provided by direct or delayed simulation.
Therefore, a technique based on fair simulation can remove more states than based direct or delayed simulation
(compare to \figref{fair_nodelayed1}). However, in contrast to the other types, merging two fair-simulation
equivalent states is not always possible without changing the language of the automaton (see \figref{fair_fail}).
We present a method which checks if merging two states changes the language with acceptable overhead.

Moreover, the presented algorithm extends techniques of \libref{simulation_general} by also removing some
transitions redundant for the language of the automaton, as seen in \figref{fair_nodelayed2}.
This approach effectively eliminates entire parts of an automaton which
become unreachable after the edge removal.\\\\
Besides presenting the algorithm, another focus of this work is to illustrate the theory behind it
and to cover optimizations useful in practice. Further we give experimental results as the
algorithm runs in the context of the ULTIMATE Project \libref{ultimate}, which is a program analysis framework.
Additionally we compare our method to other minimization techniques.

% Simulation
\section{Simulation}
This section presents different types of simulation based on relations, focusing on fair simulation.
It first demonstrates how to obtain simulation relations on Büchi automata in general.
As an introduction to the minimization algorithm, \textit{parity games} and \textit{parity game graphs}
are then explained in more detail.

% Preliminaries
\subsection{Preliminaries}
\begin{mydef}
	A \textnormal{Büchi automaton} $A$ is a tuple $\langle\Sigma, Q , Q_0, \Delta, F\rangle$, where $\Sigma$ is a finite
	set called the alphabet, $Q$ is a finite set of states, $Q_0 \subseteq Q$ is the set of initial states,
	$\Delta \subseteq Q \times \Sigma \times Q$ is the transition relation and $F \subseteq Q$ is the set of final states.
\end{mydef}\quad\\
We define $succ(v)$ for the set of successors $\{v' \in V | \exists a \in \Sigma \, \exists (v, a, v') \in \Delta\}$
and $pred(v)$ analogue for the set of predecessors of a given state $v$.

Given a Büchi automaton $A$, $A^q$ refers to the Büchi automaton\\
$\langle\Sigma, Q, \{q\}, \Delta, F\rangle$
which is the automaton $A$ with only $q$ as initial state.\\
\begin{mydef}
	A \textnormal{run} of a Büchi automaton $A$ is a sequence\\
	$\pi = q_0a_1q_1a_2q_2a_3q_3 \ldots$ of arbitrary states
	alternating with arbitrary letters such that $\forall i : (q_i, a_{i+1}, q_{i+1}) \in \Delta$.
	The \textnormal{run} forms the path $q_0 \overset{a_1}{\rightarrow} q_1 \overset{a_2}{\rightarrow}
	q_2 \overset{a_3}{\rightarrow} q_3 \ldots$ in $A$ and the corresponding $\omega$-word is
	$w = a_1a_2a_3 \ldots$.
\end{mydef}\quad\\
We write $q \in \pi$ or $a \in \pi$ if the state or letter occurs at least once in the sequence $\pi$.

The automaton $A$ accepts an $\omega$-word $w = a_1a_2a_3 \ldots$ iff for at least one corresponding \textit{run} $\pi$
the following holds:
\begin{align*}
	q_0 \in Q_0, \forall i: (q_i, a_{i + 1}, q_{i + 1}) \in \Delta \text{ and } |\{i : q_i \in F\}| = \infty,
\end{align*}
i.e. $\pi$ starts at an initial state and visits finite states infinitely often.\\
\begin{mydef}
	The accepted $\omega$-language of a Büchi automaton $A$ is\\
	$\mathcal{L}(A) = \{w \in \Sigma^\omega | A \text{ accepts } w\}$.
\end{mydef}\quad\\
For the sake of simplicity, the presented algorithm requires $A$ to have no \textit{dead ends}.\\
\begin{mydef}\label{def_deadend}
	A \textnormal{dead end} is a state of an automaton $A$ that has no outgoing transitions, i.e.
	$v$ \textnormal{dead end} $\Leftrightarrow v \in Q \land succ(v) = \emptyset$.\\
\end{mydef}
% Figure
\begin{figure}[ht]
	 \begin{center}
		\includegraphics[scale=0.5]{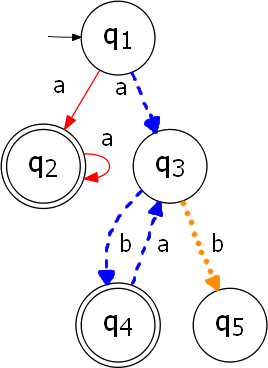}
	\end{center}
	\caption{Solid lines represent the run $\pi_1 = q_1aq_2a\ldots$, dashed lines
		the run $\pi_2 = q_1aq_3bq_4aq_3\ldots$. They form the accepting $\omega$-words
		$w_1 = aa^\omega$ and $w_2 = a(ba)^\omega$.
		The dotted transition indicates the run $\pi_3 = q_1aq_3bq_5$. This run does not build
		an accepting word because $w_3 = ab$ does not visit final states infinitely often.
		The accepted $\omega$-language of the automaton is $\mathcal{L}(A) = \{w_1, w_2\}$.
		And $q_5$ is a dead end since it has no successors.}
	\label{run_language}
\end{figure}
However, \sectionref{section_optimization} shows how automata with dead ends are handled.
\figref{run_language} shows an automaton and its language, and illustrates several
runs with their corresponding words and a dead state.

\newpage

% Types of simulation
\subsection{Types of simulation}
\begin{mydef}\label{def_simulation}\quad\\*
	\begin{enumerate}
		\item Given a Büchi automaton $A$, \textnormal{fair simulation} is defined as a relation
			$\preceq_{f} \, \subseteq Q \times Q$ where
			\begin{align*}
				q \preceq_{f} q' \text{ iff } & \left(\forall w = a_1a_2 \ldots \exists \pi = qa_1q_1a_2 \ldots
						\Rightarrow \exists \pi' = q'a_1 q'_1a_2 \ldots\right)\\*
					& \land \left(\forall w = a_1a_2 \ldots \in \mathcal{L}(A^q) \Rightarrow w \in \mathcal{L}(A^{q'})\right).
			\end{align*}
			I.e. for all words that have a corresponding \textnormal{run} starting at $q$, there
			must also be a corresponding \textnormal{run} to the same word starting at $q'$.
			And, for all accepting words whose \textnormal{run}s start at $q$, there must also exist an
			accepting \textnormal{run} starting at $q'$ that corresponds to the same word.
		\item Let $\pi = q_0a_1q_1a_2q_2a_3q_3 \ldots$ and $\pi' = q'_0a_1q'_1a_2q'_2a_3q'_3 \ldots$
			be corresponding \textnormal{run}s to $w$ starting at $q = q_0$ and $q' = q'_0$.
			Then \textnormal{delayed simulation} is a relation $\preceq_{de} \, \subseteq Q \times Q$ where
			\begin{align*}
				q \preceq_{de} q' \text{ iff } q \preceq_{f} q' \land \forall i : q_i \in F \Rightarrow \exists j \ge i : q'_j \in F.
			\end{align*}
			Every time $\pi$ visits an accepting state $q_i$, $\pi'$ must also visit at least one accepting state $q'_j$
			at some point after to cover that event.
		\item \textnormal{Direct simulation} is defined analogously,
			\begin{align*}
				q \preceq_{di} q' \text{ iff } q \preceq_{f} q' \land \forall i : q_i \in F \Rightarrow q'_i \in F.
			\end{align*}
			That is, every time $\pi$ visits an accepting state $q_i$, $\pi'$ must also visit an accepting state
			$q'_i$ at the same time.
	\end{enumerate}
\end{mydef}\quad\\
We say $q'$ $\star$-simulates $q$ if the relation $q \preceq_{\star} q'$, where $\star \in \{f, de, di\}$, holds.
Note that $q \preceq_{di} q' \Rightarrow q \preceq_{de} q' \Rightarrow q \preceq_{f} q'$, but not vice versa
which follows directly from the definition.

Again taking a look at \figref{fair_nodelayed1} reveals that $q_{l1}$ \textit{fairly simulates} $q_{r1}$
($q_{r1} \preceq_f q_{l1}$) and vice versa.
The run starting at the state $q_{r1}$ is $\pi_{r1} = q_{r1}cq_{r2}cq_{r3}bq_bc \ldots$, which corresponds
to the word $w = ccbc \ldots$. The run $\pi_{r1}$ is accepting, but there is also an accepting run
starting at $q_{l1}$ corresponding to the same word, namely $\pi_{l1} = q_{l1}cq_{l2}cq_{l3}bq_bc \ldots$.
Since there are no other accepting runs starting at $q_{r1}$ it follows that $q_{r1} \preceq_f q_{l1}$.\\\\
It holds that $q_{r1}$ \textit{delayedly} and \textit{directly simulates} $q_{l1}$. Regardless of whether $\pi_{l1}$
visits $q_a$ or $q_b$, everytime it visits an accepting state $\pi_{r1}$ does also. This is because $q_b$ is the only accepting
state on run $\pi_{l1}$.

But $q_{l1}$ does not \textit{directly simulate} $q_{r1}$. Since $q_{l1}, q_{l2}, q_{l3}$ are not accepting states,
$\pi_{l1}$ cannot visit accepting states the same time $\pi_{r1}$ does.

$q_{l1}$ also not \textit{delayedly simulates} $q_{r1}$ because the corresponding word
$w = ccac \ldots$ of the run $\pi_{r1} = q_{r1}cq_{r2}cq_{r3}aq_ac \ldots$
can only be matched by the run $\pi_{l1} = q_{l1}cq_{l2}cq_{l3}aq_ac \ldots$.
Further, $\pi_{r1}$ visits three accepting states before entering the loop whereas
$\pi_{l1}$ not visits an accepting state for coverage.\\\\
Last we point out why $q_{r1}$ does not \textit{fairly simulate} $q_{r2}$.
The run $\pi_{r2} = q_{r2}cq_{r3}bq_bc \ldots$ corresponds to the word $w = cbc \ldots$,
but no run $\pi_{r1}$ corresponding to the same word starting from $q_{r1}$ exists.\\\\
For minimizing automata pairs of states that simulate each other, i.e. $q$ and
$q'$ if $q \preceq_{\star} q' \land q' \preceq_{\star} q$ are of interest. As seen later, for $\star \in \{di, de\}$ it is
always possible to merge such a pair without changing the language of the automaton.
For fair simulation this is not always be the case. As the algorithm in question aims to
reduce as many states as safely possible, the correctness of any such potential merge
must be verified before it is applied.\\\\
A naive implementation, for determining whether a pair of states $q$ and $q'$ simulates each other, consists
of iterating over all possible words starting at $q$ and $q'$. Then creating all corresponding runs for such
words and compute wether \defref{def_simulation} holds.

We present a more efficient implementation, \algoref{algo_jurdzinski}.
The concept of this algorithm is based on \textit{parity games}.

% Parity game
\subsection{Parity game}
A parity game $G_A(q, q')$, where $q$ and $q'$ are arbitrary states of the Büchi automaton $A$, is played by two players,
\textit{Spoiler} (or \textit{Antagonist}) and \textit{Duplicator} (or \textit{Protagonist}). Each player has one token,
initially placed at states $q$ and $q'$, for \textit{Spoiler} and \textit{Duplicator} respectively.
The players move their token alongside the automaton $A$ using transitions. Both tokens can be moved to the same state
without interfering each other.

The game is played in rounds, assuming that at the beginning of round $i$ Spoiler's token is at state $q_i$
and Duplicator's at $q'_i$, a round is played as follows:
\begin{enumerate}
	\item Spoiler chooses a transition $(q_i, a, q_{i + 1}) \in \Delta$ and moves his token to $q_{i + 1}$.
	\item Duplicator must now respond to Spoiler's choice, choose an \textit{a-transition} $(q'_i, a, q'_{i + 1}) \in \Delta$ and move
		his token to $q'_{i + 1}$.
\end{enumerate}
If $q'_i$ does not have such an outgoing a-transition, Duplicator can not respond.
The game halts and Spoiler wins the game.
Note that Spoiler can always choose an outgoing transition, as we assume
$A$ having no \textit{dead ends} (compare to \defref{def_deadend}).

If Duplicator can always respond to Spoilers choices, the game produces two infinite \textit{run}s
$\pi = qa_1q_1a_2q_2a_3q_3 \ldots$ and $\pi' = q'a_1q'_1a_2q'_2a_3q'_3 \ldots$.
The \textit{game history} is defined as the tuple $(\pi, \pi')$.\\\\
We connsider three types of parity games. All of them differ in the winning conditions for Duplicator.

\begin{mydef}\label{def_gamewinning}\quad\\*
	\begin{enumerate}
		\item In the \textnormal{direct simulation game} $G^{di}_A(q, q')$ the game history $(\pi, \pi')$ is winning for Duplicator
			iff $\forall i : q_i \in F \Rightarrow q'_i \in F$.
		\item In the \textnormal{delayed simulation game} $G^{de}_A(q, q')$ the game history $(\pi, \pi')$ is winning for Duplicator
			iff $\forall i : q_i \in F \Rightarrow \exists j > i : q'_j \in F$.
		\item In the \textnormal{fair simulation game} $G^f_A(q, q')$ the game history $(\pi, \pi')$ is winning for Duplicator
			iff $|\{j | q'_j \in F\}| = \infty \Rightarrow |\{i | q_i \in F\}| < \infty$.
	\end{enumerate}
\end{mydef}\quad\\
Note that there are similarities between this definition and the definition of simulation (\defref{def_simulation}).\\
% Figure
\begin{figure}[ht]
	 \begin{center}
		\includegraphics[scale=0.6]{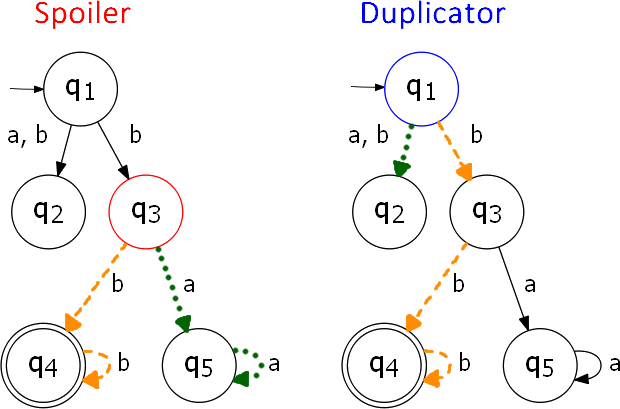}
	\end{center}
	\caption{A parity game with Spoiler starting at $q_3$ and Duplicator at $q_1$. The dashed and dotted lines represent
		the two possible runs Spoiler can create and possible responses of Duplicator.}
	\label{paritygame}
\end{figure}\quad\\
\figref{paritygame} shows two copies of the same automata. They represent a parity game with the token of Spoiler
initially standing on $q_3$ and the token of Duplicator on $q_1$. Spoiler has two possibilites in the first round, he can choose the
$b$-transition $(q_3, b, q_4)$ or the $a$-transition $(q_3, a, q_5)$. A wise decision would be to choose the $a$-transition,
we assume he decides for this way. The token of Spoiler now is placed at $q_5$ and Duplicator must
also choose an $a$-transition or he looses. Duplicator decides for $(q_1, a, q_2)$ and moves his token to $q_2$.
The second round starts and Spoiler has again the option to choose any transition he likes.
However, there is only one transition and he chooses $(q_5, a, q_5)$. Duplicator must react and also choose an
$a$-transition but $q_2$ has no outgoing transitions. The game holds and Spoiler wins since Duplicator
can not match his transition.

Now assuming Spoiler takes the $b$-transition in the first round and moves to $q_4$ instead of $q_5$.
If Duplicator takes the transition $(q_1, b, q_2)$ he would lose again in the next round so he chooses $(q_1, b, q_3)$.
The only possibility for Spoiler is to take $(q_4, b, q_4)$ and Duplicator also moves to $q_4$. Both players now
repeatedly take the transition $(q_4, b, q_4)$ and the game creates the game history $(\pi, \pi')$
where $\pi = q_3bq_4b \ldots$ and $\pi' = q_1bq_3bq_4b \ldots$. The game history determines the winner,
for \textit{fair} and \textit{delayed simulation} the game history is winning for Duplicator, for \textit{direct simulation}
Spoiler wins since he visits an accepting state in the first round while Duplicator visits $q_3$ in this round.

Observe that as soon as, in the beginning of the same round, Duplicator's token is placed
at the same state Spoiler's token is at, he can directly copy every move of Spoiler by simply following him.
If that is the case Spoiler cannot win the game anymore.\\\\
While assuming both players give their best Spoiler would not choose the
transition $(q_3, b, q_4)$ and Duplicator would not choose $(q_1, b, q_2)$.

% Strategy
\subsubsection{Strategy}
Informally a \textit{stategy} for Duplicator determines at each round of the game which transition Duplicator should
choose based on the history of previous rounds. Such a strategy is called a \textit{winning strategy} if, no matter
how Spoiler plays, Duplicator always wins. While assuming that Duplicator always gives his best to win this
corresponds to using the best strategy. Formally \textit{strategy} and \textit{winning strategy}
are defined the way as seen in \libref{simulation_general}.\\
\begin{mydef}\label{def_strategy}
	Let $s_i = q_0q'_0a_1q_1q'_1a_2 \ldots a_{i-1}q_{i-1}q'_{i-1}a_iq_i$ be an
	\textnormal{interleaving sequence} such that
	$\pi' = q_0a_1q_1a_2 \ldots a_{i-1}q_{i-1}a_iq_i$ is a run for Spoiler and
	$\pi = q'_0a_1q'_1a_2 \ldots a_{i-1}q'_{i-1}$ is a run for Duplicator.
	
	A \textnormal{strategy} for Duplicator in $G^{\star}_A(q, q')$ is a partial function
	$f : Q \times (Q \times \Sigma \times Q)^* \pto Q$ where
	\begin{align*}
		f(q) = q' \land (\forall s_i : i > 0 \Rightarrow (q'_{i-1}, a_i, f(s_i)) \in \Delta).
	\end{align*}
\end{mydef}\quad\\
Observe that the existance of a \textit{strategy} implies that it is possible for Duplicator
to play the game in such way that it does not halt. It also means that if there does not exist
a \textit{strategy} for Duplicator then Spoiler always wins the game since he can
make moves Duplicator cannot respond to.

In the example of \figref{paritygame} there does not exist a \textit{strategy} for Duplicator. Since if Spoiler
chooses the $a$-transition $(q_3, a, q_5)$ and then $(q_5, a, q_5)$ Duplicator, meanwhile at $q_2$,
has no possibility to find an outgoing $a$-transition.
The corresponding sequence up to this point is $s_2 = q_3q_1aq_5q_2aq_5$.

Assuming the transition $(q_3, a, q_5)$ does not exist, Spoiler must choose the $b$-transition $(q_3, b, q_4)$,
then there exist two different strategies. The first strategy $f_1$ chooses $(q_1, b, q_2)$ for
Duplicator, $f_1(q_3q_1bq_4) = q_2$ and the second strategy $f_2$
chooses $(q_1, b, q_3)$, $f_2(q_3q_1bq_4) = q_3$.\\
\begin{mydef}
	A strategy $f$ for Duplicator in $G^{\star}_A(q, q')$ is a \textnormal{winning strategy} iff
	for all runs of Spoiler $\pi$ there exists a run of Duplicator $\pi'$, received by using the strategy for each move,
	such that the game history $(\pi, \pi')$ is winning for Duplicator \textnormal{(\defref{def_gamewinning})}.
	
	$\forall \pi = q_0a_1q_1a_2 \ldots \exists \pi' : (\pi, \pi')$
	winning for Duplicator, where run $\pi' = q_0'a_1q_1'a_2 \ldots$ is the run received by
	$q'_{i+1} = f(q_0q'_0a_1q_1q'_1a_2 \ldots q_{i+1})$.
\end{mydef}\quad\\
Having a winning strategy means that, for fixed starting positions, no matter how Spoiler plays the
strategy will always create a winning game history for Duplicator. When using such a strategy for given
starting positions it is not possible for Spoiler to win the game.

In \figref{paritygame}, when again assuming $(q_3, a, q_5)$ does not exist, there is a winning strategy
for \textit{fair} and \textit{delayed simulation} games. Strategy $f_2$, which chooses $(q_1, b, q_3)$ instead of $(q_1, b, q_2)$
is a winning strategy. In complete $f_2$ is defined as follows:
\begin{align*}
	f_2	&(q_3)				&= q_1\\
	f_2	&(q_3q_1bq_4)			&= q_3\\
	f_2	&(q_3q_1bq_4q_3bq_4)		&= q_4\\
	f_2	&(q_3q_1bq_4q_3bq_4q_4bq_4)	&= q_4\\
		&\vdots
\end{align*}\\\\
The following lemma shows the connection between parity games and simulation relations.
\begin{mylemma}\label{lemma_parity_game}
	Given a Büchi automaton $A$ where $q$ and $q'$ are states of $A$,\\
	$q \preceq_{\star} q'$ iff there exists a winning strategy for Duplicator in $G^{\star}_A(q, q')$ where $\star \in \{di, de, f\}$.
\end{mylemma}
\begin{proof}
	The \textit{winning strategy} creates game histories $(\pi, \pi')$ for every run $\pi$ Spoiler can build.
	Since the strategy is a \textit{winning strategy} every run $\pi'$ that follows the strategy creates a game
	history that is winning for Duplicator. Given the definition of such a game history
	(\defref{def_gamewinning}) it follows directly that $q \preceq_{\star} q'$.
	
	Assuming $q \preceq_{\star} q'$ there must exist a game history $(\pi, \pi')$ that is winning for Duplicator for
	every possible run $\pi$ Spoiler can create.
	All possible runs $\pi$ of Spoiler together with the corresponding runs $\pi'$ of Duplicator define
	a \textit{winning strategy} and the proof is complete.\\
\end{proof}\quad\\
Given two states $q$ and $q'$, finding out if $q \preceq_{\star} q'$ is equivalent to finding a winning
strategy for Duplicator in the game $G^{\star}_A(q, q')$.

% Parity game graph
\subsection{Parity game graph}
Obviously the difficulty is to find a winning strategy or to proof that there does not exist one.
By using parity game graphs the problem can be solved algorithmic.
\begin{mydef}\label{def_paritygamegraph}
	Let $A = \langle\Sigma, Q , Q_0, \Delta, F\rangle$ be a Büchi automaton to which we refer
	as Spoiler's automaton and $A' = \langle\Sigma, Q' , Q'_0, \Delta', F'\rangle$ is a Büchi automaton
	to which we refer as Duplicator's automaton.
	A \textnormal{parity game graph} on two Büchi automata is a tuple
	$G^{\star}_{A, A'} = \langle V_0^{\star}, V_1^{\star}, E^{\star}, p^{\star}\rangle$
	where $\star \in \{di, de, f\}$.
	
	$V^{\star}$ is the set of vertices and $V^{\star} = V_0^{\star} \cup V_1^{\star}$, $E^{\star} \subseteq V^{\star} \times V^{\star}$
	is the set of edges (note that labels for the edges are not needed) and $p^{\star}: V^{\star} \to \mathbb{N}$ is
	a function that maps a priority to each vertex.
	
	The elements are defined as follows:
	\begin{enumerate}
		\item For fair simulation ($\star = f$):
			\subitem $V_1^f = \{v_{(q,q')} | q \in Q \land q' \in Q'\}$
			\subitem $V_0^f = \{v_{(q, q', a)} | q \in Q \land q' \in Q' \land \exists \tilde{q} \in pred(q) : (\tilde{q}, a, q) \in \Delta\}$
			\subitem $E^f = \underbrace{\{(v_{(q, q')}, v_{(\tilde{q}, q', a)}) | \exists (q, a, \tilde{q}) \in \Delta\}}_{\text{move from Spoiler}}$
				\subsubitem $\cup \underbrace{\{(v_{(q, q', a)}, v_{(q, \tilde{q})}) |
					\exists (q', a, \tilde{q}) \in \Delta'\}}_{\text{move from Duplicator}}$
			\subitem $p^f: V^f \to \{0, 1, 2\}, p^f(v) = \begin{cases} 0 & \textnormal{if } v = v_{(q, q')} \land q' \in F'\\
				1 & \textnormal{if } v = v_{(q, q')} \land q \in F \land q' \notin F'\\
				2 & \textnormal{otherwise}
				\end{cases}$
		\item For direct simulation ($\star = di$):
			\subitem $V_1^{di} = V_1^f$
			\subitem $V_0^{di} = V_0^f$
			\subitem $E^{di} = E^f \setminus (\{(v_{(q, q')}, v_{(\tilde{q}, q', a)}) | q \in F \land q' \notin F'\}$
				\subsubitem $\cup \{(v_{(q, q', a)}, v_{(q, \tilde{q})}) | q \in F \land \tilde{q} \notin F'\})$
			\subitem $p^{di}: V^{di} \to \{0\}, p^{di}(v) = 0$
		\item For delayed simulation ($\star = de$):
			\subitem $V_1^{de} = \{v_{(b, q, q')} | q \in Q \land q' \in Q' \land b \in \{0, 1\} \land (q' \in F' \rightarrow b = 0)\}$
			\subitem $V_0^{de} = \{v_{(b, q, q', a)} | q \in Q \land q' \in Q' \land b \in \{0, 1\} \land \exists \tilde{q} \in pred(q) : (\tilde{q}, a, q) \in \Delta\}$
			\subitem $E^{de} = \{v_{(b, q, q'), v_{(b, \tilde{q}, q', a)}} | (q, a, \tilde{q}) \in \Delta \land \tilde{q} \notin F\}$
				\subsubitem $\cup \{v_{(b, q, q'), v_{(1, \tilde{q}, q', a)}} | (q, a, \tilde{q}) \in \Delta \land \tilde{q} \in F\}$
				\subsubitem $\cup \{v_{(b, q, q', a)}, v_{(b, q, \tilde{q})} | (q', a, \tilde{q} \in \Delta' \land \tilde{q} \notin F')\}$
				\subsubitem $\cup \{v_{(b, q, q', a)}, v_{(0, q, \tilde{q})} | (q', a, \tilde{q} \in \Delta' \land \tilde{q} \in F')\}$
			\subitem $p^{de}: V^{de} \to \{0, 1, 2\}, p^{de}(v) = \begin{cases} b & \textnormal{if } v = v_{(b, q, q')} \in V^{de}_1\\
				2 & \textnormal{if } v \in V^{de}_0
				\end{cases}$
		\item[] The bit \textnormal{b} encodes whether Spoiler has visited a final state
			without Duplicator visiting one since then, 1 indicates this and 0 the opposite case.
	\end{enumerate}
	For the remainder of this thesis $(q, q')$ abbreviates the vertex $v_{(q, q')}$ and $(q, q', a)$ the vertex $v_{(q, q', a)}$.
\end{mydef}\quad\\
% Figure
\begin{figure}[ht]
	 \begin{center}
		\includegraphics[scale=0.7]{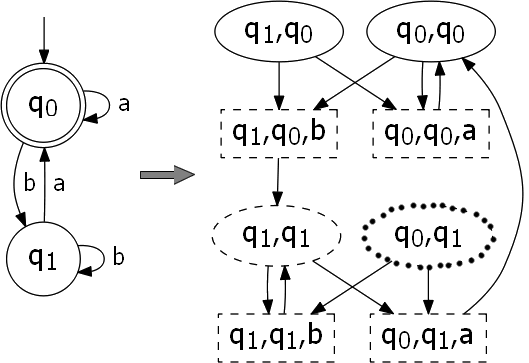}
	\end{center}
	\caption{The automaton $A$ from \figref{fair_fail} and its parity game graph for fair simulation $G^f_{A, A}$.
		The elliptical shaped vertices are in $V_1^f$ and the box shaped in $V_0^f$. The vertices with a solid
		border have a priority $p$ of $0$, the dashed have priority $2$ and the dotted
		vertex $v_{(q_0, q_1)}$ has priority $1$.}
	\label{paritygamegraph}
\end{figure}\quad\\
From now the focus is on fair simulation parity game graphs but the same can be applied similiar to other simulations.

First note that a parity game graph is build using two automata $A$ and $A'$. In order to compute simulation
relations it is enough to use $A = A'$ and simply write $G^f_A$ instead of $G^f_{A, A}$.
But \sectionref{section_minimization} needs the possibility to build a game graph on two different automata.
This allows letting Spoiler play on a different automaton than Duplicator which makes computing simulation
relations between two different automaton possible.

A parity game graph represents all possible positions and moves of a parity game in one graph.
For each position of the parity game, the states Spoiler and Duplicator stand on,
there is a vertex in the graph. Possible moves of the players are represented by edges
between the vertices.

More precisely a vertex $v_1 = (q, q') \in V^f_1$ or $v_0 = (q, q', a) \in V^f_0$
encodes the position where Spoiler's token is placed at $q$ and Duplicator's at $q'$.
For such a vertex $v_1$ it is now Spoiler's time to make a move and choose a transition, he may choose
an a-transition $(q, a, \tilde{q}) \in \Delta$ which is encoded by the vertex $v_0 = (\tilde{q}, q', a) \in V^f_0$
that stands for the position where Duplicator also needs to choose an a-transition. Let that transition
be $(q', a, \bar{q}) \in \Delta'$ which then leads to the vertex $(\tilde{q}, \bar{q}) \in V^f_1$ in the game graph.
The game graph has edges between vertices for every possible move. For example the edge
$(v_{(q, q')}, v_{(\tilde{q}, q', a)})$ does exist if it is possible for Spoiler to move from the state
$q$ to $\tilde{q}$ by using the letter $a$, i.e. $(q, a, \tilde{q}) \in \Delta$. And $(v_{(q, q', a)}, v_{(q, \tilde{q})})$
exists if Duplicator can move from $q'$ to $\tilde{q}$ by using $a$, i.e. $(q', a, \tilde{q}) \in \Delta'$.

\figref{paritygamegraph} illustrates the fair simulation game graph $G^f_{A, A}$ obtained by using
the automaton $A$ (from \figref{fair_fail}) for Spoiler and also for Duplicator. When creating a game graph,
the transition $(q_0, a, q_0)$ from $A$ produces the edges $(v_{(q_0, q_0)}, v_{(q_0, q_0, a)})$,
$(v_{(q_0, q_1)}, v_{(q_0, q_1, a)})$ and $(v_{(q_0, q_0, a)}, v_{(q_0, q_0)})$ in $G^f_{A, A}$.
The vertex $v_{(q_1, q_0, a)}$ does not exist since it would require $q_1$
to have an incoming transition labeled with $a$.

% Correlation to parity games
\subsubsection{Correlation to parity games}\label{section_gamegraphtogame}
A parity game $G^f_A(q, q')$ can be played on the corresponding parity game graph by starting at the vertex $(q, q') \in V_1$.
The two players, with Spoiler starting, now alternately move the same token over the game graph representing the original game.

The priorities encode the simulation conditions. Priority $0$, for fair simulation, represents the situation where Duplicator visits a
final state. Analogously, priority $1$ stands for the position where Spoiler visits a final state and Duplicator does not.
Priority $2$ is a neutral situation in which both players do not visit a final state.

$\figref{paritygamegraph}$ demonstrates the definition of priority for \textit{fair simulation}. The vertex $v_{(q_1, q_1)}$
has a priority of $2$ since $q_1$ is no final state, all vertices in $V_0$ also have priority $2$.
However, the vertex $v_{(q_0, q_1)}$ has priority $1$ because Spoiler's state
$q_0$ is final while Duplicator's state $q_1$ is not.\\\\
\defref{def_gamewinning} describes that Duplicator wins a parity game if he does not allow Spoiler to visit
final states infinitely often, while not doing the same. When Duplicator lost, then he visited vertices with
priority $1$ infinitely often by not also visiting priority $0$ that many.
So in general Duplicator prefers to visit vertices with priority $0$ and Spoiler prefers priority $1$ since those priorities bring
the players closer to their victory. Although this sounds like a good strategy for Duplicator, if he dully moves to a vertex $v$
with priority $0$, this could lead him into a trap. For example a loop of vertices with priority $1$ behind $v$ or similar.\\\\
The following two definitions clarify the connection between parity games and game graphs
by introducing an isomorphism $\sigma$.
\begin{mydef}\label{def_isomorphism}
	Given the Büchi automata $A, A'$ and the game graph $G^f_{A, A'}$, let $P$
	be the set of all \textnormal{paths} in $G^f_{A, A'}$, $R$ the set of all \textnormal{runs} in
	$A$ and $R'$ of all \textnormal{runs} in $A'$.
	
	Then we define the \textnormal{path-transformation} $\sigma: P \ito (R \times R')$.
	Which is, for a path $\varrho = v_{(q_0, q'_0)} \rightarrow v_{(q_1, q'_0, a_1)}
	\rightarrow v_{(q_1, q'_1)} \rightarrow \ldots$, given by
	$\sigma(\varrho) = (\pi, \pi')$. Where $\pi = q_0a_1q_1 \ldots$ and $\pi' = q'_0a_1q'_1 \ldots$.
\end{mydef}
\begin{mylemma}
	The \textnormal{path-transformation} $\sigma$ is an \textnormal{isomorphism} between
	paths in game graphs and game histories in parity games.
\end{mylemma}
\begin{proof}
	$\sigma$ is a morphism by definition. We define $\sigma^{-1} : (R \times R') \ito P$ given
	by $\sigma^{-1}((\pi, \pi')) = \varrho$ where $\pi, \pi'$ and $\varrho$ are defined the
	same as in \defref{def_isomorphism}. It follows that $\sigma \circ \sigma^{-1} = \operatorname{id}_{R \times R}$
	and $\sigma^{-1} \circ \sigma = \operatorname{id}_P$. Thus $\sigma$ is bijective, an ismorphism and $\sigma^{-1}$
	is the inverse of $\sigma$.\\
\end{proof}\quad\\
Given the isomorphism $\sigma$, we define when a path is \textit{winning} for Duplicator.
\begin{mydef}\label{def_pathwinning}
	A path $\varrho$ in a game graph $G^f_{A, A'}$ is \textnormal{winning} for Duplicator
	iff the game history $\sigma(\varrho)$ is \textnormal{winning} for Duplicator.
\end{mydef}\quad\\
For simplicity $v \in \varrho$ means the path visits a vertex $v$.
With the aid of vertex priorities, the following lemma enhances \defref{def_pathwinning}.
\begin{mylemma}\label{lemma_pathwinningext}
	The \textnormal{path} $\varrho$ is \textnormal{winning} for Spoiler iff
	$min\{n : |\{v | v \in \varrho \land p^f(v) = n\}| = \infty\}$ is \textnormal{odd},
	i.e. the smallest priority that occurs infinite times in the path.
\end{mylemma}
\begin{proof}
	Spoiler only wins if he visits final states infinitely times while Duplicator does not. This case is reflected by
	a path that visits vertices with priority $1$ infinitely times while not visiting priority $0$ infinitely often.
	In all other cases Duplicator wins the game. Since $1$ is the only odd priority, the lemma holds.\\
\end{proof}\quad\\
Using the path-transformation $\sigma$ and \lemmaref{lemma_pathwinningext}, a \textit{winning strategy}
can be obtained on a parity game graph $G^f_{A, A'}$ only.

Let us deepen that by again taking a look on \figref{paritygamegraph}.
When playing a parity game on the game graph $G^f_A$, starting at $(q_1, q_1)$,
Spoiler may move to $(q_1, q_1, b)$ or $(q_0, q_1, a)$. Assuming he decided for $(q_1, q_1, b)$, Duplicator can
only choose $(q_1, q_1)$ and it is Spoiler's turn, again on the same spot. He may choose this edge everytime
which produces the \textit{path} $\varrho = v_{(q_1, q_1)} \rightarrow v_{(q_1, q_1, b)} \rightarrow v_{(q_1, q_1)} \rightarrow \ldots$.
The smallest priority that occurs infinite times on $\varrho$ is $2$, even both vertices have that priority, which is not odd.
Because of that $\varrho$ is winning for Duplicator but Spoiler may choose the transition to $(q_0, q_1, a)$ and so on.

In this example, it is not possible for Spoiler to win the play since there is no
possibility to visit priority $1$ infinite times no matter where the starting position is or how Duplicator plays.
For this reason every strategy for Duplicator is a winning strategy in this game.\\\\
The correlation between game graphs and parity games leads to the next observation.
With \lemmaref{lemma_parity_game}, $\sigma$ and \lemmaref{lemma_pathwinningext}, a simulation
relation can be computed on a parity game graph $G^f_{A, A'}$ only.\\\\
Referring to the example of \figref{paritygamegraph}, where every strategy defines a winning strategy,
the fair simulation relation is obtained. The relation consists of $q_0 \preceq_f q_1$ and $q_1 \preceq_f q_0$.
As seen before such a pair that \textit{fairly simulates} each other is a candidate for merging.
However, as seen later, merging $q_0$ and $q_1$ will change the language of the
automaton from $\{w \in \Sigma^{\omega} | w = (b^*a)^{\omega}\}$ to
$\{w \in \Sigma^{\omega} | w = (ab)^{\omega}\}$ so the merge attempt gets rejected.

% Computing a simulation relation
\section{Computing a simulation relation}
In the section before we have seen that $q \preceq_f q'$ if there exists a winning strategy for Duplicator
in the parity game $G^f_A(q, q')$ (see \lemmaref{lemma_parity_game}).
This section shows how to compute if there exists a winning strategy for a pair of states $q$ and $q'$.
A winning strategy is computed on the game graph only by using the isomorphism
$\sigma$ (\defref{def_isomorphism}) and \lemmaref{lemma_pathwinningext}.

With the use of \textit{Jurdzi\'nskis algorithm}, originally from \libref{jurdz_original}, the simulation
relation is computed for every pair of starting states at the same time. This is done by building paths on the game graph
that reflect the optimal solution for the players and extending them in every round if possible.
The paths are computed by introducing a progress measure for each vertex.\\\\
Given a parity game graph $G^f_{A, A'} = \langle V_0, V_1, E, p\rangle$ on two Büchi automata $A$ and $A'$,
in order for the algorithm to work correctly we assume that the graph has no self loops nor dead ends (\defref{def_deadend}).
The first condition will not occur if the graph was built correctly but dead ends can develop in practice.
However, \sectionref{section_optimization} shows how to handle these.

Note that \textit{Jurdzi\'nskis algorithm} also works more general on priorities up to an arbitrary $k \in \mathbb{N}$, as shown
in \libref{simulation_general}.

% Terminology
\subsection{Terminology}\label{section_computingterminology}
This section starts with some terminology. First the priority function for fair simulation parity game graphs
$p^f$ (\defref{def_paritygamegraph}) is needed. For simplicity $p$ abbreviates $p^f$.
\begin{mydef}
	We define $n$ as the amount of vertices that have a priority of $1$,
	$n = |\{v : v \in V \land p(v) = 1\}|$.
	
	Next $\mu$ is a function that assigns each vertex a progress measure,
	$\mu: V \to \left(\{0, 1, \ldots, n\} \cup \{\infty\}\right)$, where $\forall i \in \{0, 1, \ldots, n\}: i < \infty$.
\end{mydef}\quad\\
The algorithm uses a set of functions $\text{incr}_i$. They are used to increase the current progress measure
of a vertex based on its priority $i$.
\begin{mydef}
	Given the priority $i$ of a vertex, $i \in \{0, 1, 2\}$, we define the function
	$\text{incr}_i : (\{0, 1, \ldots, n\} \cup \{\infty\}) \to (\{0, 1, \ldots, n\} \cup \{\infty\})$ as follows.
	\begin{align*}
		\text{incr}_i(x) = \begin{cases} x + 1 & \text{if } i = 1 \land x < n\\
					x & \text{if } i = 2 \land x \neq \infty\\
					0 & \text{if } i = 0 \land x \neq \infty\\
					\infty & \text{if } x = \infty \lor (i = 1 \land x = n)
				\end{cases}
	\end{align*}
\end{mydef}\quad\\
The progress measure, together with $\text{incr}_i$, is used to count the amount of vertices visited
with priority $1$ and reset the counter if a vertex with priority $0$ is visted. The function $\text{incr}_i$
increases a progress measure from $0$ to $n$ and then to $\infty$ whenever vertices with priority
$1$ are visited. It resets the counter to $0$ if a vertex with priority $0$ gets visited
and does not change it for priority $2$.

Note that, for a fixed priority $i$, the function $\text{incr}_i(\cdot)$ is monotonically increasing.\\\\
If the progress measure of a vertex is \textit{infinity}, there does not exist a winning strategy for a
game starting at this vertex. When the algorithm terminates and a progress measure of a vertex is not
\textit{infinity}, a winning strategy for a game starting at this vertex does exist.
\sectionref{section_compsimrelation} deepens this later.

Next the algorithm needs a function that selects the \textit{best} neighboring progress measure.
It is the progress measure of a vertex, a player would choose as successor if it is his turn.
\begin{mydef}
	We define the \textnormal{best neighboring progress measure function} $\text{best-nghb-ms}:
	\{\mu | \mu: V \to (\{0, 1, \ldots, n\} \cup \{\infty\})\} \times V \to\\
	(\{0, 1, \ldots, n\} \cup \{\infty\})$, given by
	\begin{align*}
		\text{best-nghb-ms}(\mu, v) = \begin{cases} 0 & \text{if } p(v) = 0\\
								min\left(\{\mu(w) | w \in succ(v)\}\right) & \text{if } p(v) \neq 0 \land v \in V_0\\
								max\left(\{\mu(w) | w \in succ(v)\}\right) & \text{if } p(v) \neq 0 \land v \in V_1
							\end{cases}
	\end{align*}
\end{mydef}\quad\\
If its Duplicator's turn he will choose the neighbor with the smallest measure,
analogue the greatest measure for Spoiler.
If the priority of the current vertex is $0$ however, it represents a vertex that resets the progress measure counter anyways
so every choice will be optimal for both players.\\\\
A very simple although not fast implementation is to initiate every vertex with progress measure $0$
and run $\text{incr}_{p(v)}(\text{best-nghb-ms}(\mu, v))$ on each vertex until there is no progress. For all vertices $v$ that then
have a progress measure $\mu(v) \neq \infty$, there does exist a winning strategy in the game starting at $v$.
This is exactly what \algoref{algo_jurdzinski_original}, the original version of \textit{Jurdzi\'nskis algorithm} does.
For improving the runtime of the algorithm additional fields are needed.

Let $B$ and $C$ be arrays. $B$ stores the value of $\text{best-nghb-ms}(\mu, v)$ for each vertex $v$.
\begin{mydef}
	We define the \textnormal{neighbor counter function} $\text{nghb-cnt}:
	\{\mu | \mu: V \to (\{0, 1, \ldots, n\} \cup \{\infty\})\} \times V \to \mathbb{N}$,
	\begin{align*}
		\text{nghb-cnt}(\mu, v) = \begin{cases}|\{u : u \in succ(v) \land \mu(u) =\\
								\qquad\text{best-nghb-ms}(\mu, v)\}| & \text{if } p(v) \ne 0\\
							|\{u : u \in succ(v) \land 0 =\\
								\qquad\text{best-nghb-ms}(\mu, v)\}| & \text{if } p(v) = 0
						\end{cases}
	\end{align*}
\end{mydef}\quad\\
The function $\text{nghb-cnt}$ counts the number of neighbors a vertex has that represent the best choice to move at.
The array $C$ stores for each vertex $v$ the value of $\text{nghb-cnt}(\mu, v)$.

% Enhanced version of Jurdzinski algorithm
\subsection{Enhanced version of Jurdzi\'nskis algorithm}
Let us now take a brief look at \algoref{algo_jurdzinski}, which is an optimized version of the algorithm
seen in \libref{simulation_general}, and going in deep afterwards in \sectionref{section_illustration}.\\\\
% Algorithm implementation
\IncMargin{1em}
\begin{algorithm}
\SetKwData{v}{v}\SetKwData{V}{V}\SetKwData{L}{L}\SetKwData{t}{t}\SetKwData{P}{P}\SetKwData{w}{w}
\SetKwFunction{B}{B}\SetKwFunction{C}{C}\SetKwFunction{p}{p}\SetKwFunction{pm}{$\mu$}
\SetKwFunction{bestnghbms}{best-nghb-ms}\SetKwFunction{nghbcnt}{nghb-cnt}\SetKwFunction{incr}{incr}
\SetKwFunction{succ}{succ}\SetKwFunction{pred}{pred}
\For{$\v \in \V$}{
	$\B(\v) := 0$\;
	$\C(\v) := |\{\w : \w \in \succ(\v)\}|$\;
	$\pm(\v) := 0$\;
}
$\L := \{\v \in \V | \p(\v) = 1\}$
\BlankLine
\BlankLine
\While{$\L \neq \emptyset$}{
	$\text{let } \v \in \L$\;
	$\L :=\L \setminus \{\v\}$\;
	$\t := \pm(\v)$\;
	\BlankLine
	$\B(\v) := \bestnghbms(\pm, \v)$\;
	$\C(\v) := \nghbcnt(\pm, \v)$\;
	$\pm(\v) := \incr_{\p(\v)}(\B(\v))$\;
	\BlankLine
	$\P := \{\w \in \V | \w \in \pred(\v) \land \w \notin \L\}$\;
	\For{$\w \in \P$}{
		\If{$\p(w) \neq 0 \land \pm(\v) > \B(\w)$}{
			\If{$\w \in \V_1$}{
				$\L := \L \cup \{\w\}$\;
			}\ElseIf{$\w \in \V_0 \land ((\p(\w) \neq 0 \land \t = \B(\w)) \lor (\p(\w) = 0 \land 0 = \B(\w)))$}{
				\If{$\C(\w) > 1$}{
					$\C(\w) := \C(\w) - 1$\;
				}\ElseIf{$\C(\w) = 1$}{
					$\L := \L \cup \{\w\}$\;
				}
			}
		}
	}
}
\BlankLine
\caption{Efficient implementation of \textit{Jurdzi\'nskis algorithm} fitted for use with three priorities.}\label{algo_jurdzinski}
\end{algorithm}\DecMargin{1em}\quad\\
\textbf{Lines 1-5} initialize the data structures of the algorithm, every vertex gets the progress measure $0$
which also means that every neighbor $w$ has $\mu(w) = 0$ thus $C(v)$ needs to be the amount of successors.

In \textbf{line 5} a \textit{working list} is created that contains every vertex which has a priority of $1$.

\textbf{Lines 6-22} represent the loop of the algorithm that processes the working list.
Given a vertex $v$ from the working list its values are updated in \textbf{lines 10-12}.

\textbf{Lines 13-22} process the predecessors of the current vertex $v$.
If the progress measure of $v$ has increased the predecessors may be added to the working list if
they choose $v$ as optimal choice to move at.

\textbf{Lines 16-17} are responsible for the predecessors in $V_1$, they represent a move
by Spoiler and he is always interested in an increased progress measure.
\textbf{Lines 18-22} are responsible for the predecessors in $V_0$, analogously they represent a move
by Duplicator and he tries to evade the update by choosing an alternative if possible.

% Complexity
\subsubsection{Complexity}
\algoref{algo_jurdzinski} runs in $\mathcal{O}(|Q|^3 \cdot |\Delta|)$ time and
$\mathcal{O}(|Q| \cdot |\Delta|)$ space. In the following this section analyses and proofs this claim.\\\\
Given a Büchi automaton $A = \langle \Sigma, Q, Q_0, \Delta, F\rangle$ first the
size of the game graph $G^f_A$ gets analysed.\\
\begin{mylemma}\label{lemma_gamegraphcomplexity}
	For a game graph $G^f_A = \langle V_0, V_1, E, p\rangle$ with $V = V_0 \times V_1$ where
	$n = |\{v \in V : p(v) = 1\}|$ it holds that
	\begin{align*}
		|V|, |E|			&\in \mathcal{O}(|Q| \cdot |\Delta|)\\
		n	&\in \mathcal{O}(|Q|^2)
	\end{align*}
\end{mylemma}
\begin{proof}
	Looking at \defref{def_paritygamegraph} obviously $|V_1| = |Q|^2$ since for every pair $(q, q')$
	exactly one vertex gets created. Every state has at least one outgoing transition because
	a requirement of the algorithm is that $A$ has no dead ends, this follows $|Q| \le |\Delta|$.
	Furthermore $|V_1| = |Q|^2 \le |Q| \cdot |\Delta|$ thus $|V_1| \in \mathcal{O}(|Q| \cdot |\Delta|)$.

	Analyzing the size of $V_0$, every state $q'$ and transition $(q, a, \tilde{q})$
	creates a vertex $v_{(\tilde{q}, q', a)}$. This implies $|V_0| \le |Q| \cdot |\Delta|$ and the following is received:
	\begin{align*}
		|V| = |V_0| + |V_1| \le 2 \cdot (|Q| \cdot |\Delta|) \Rightarrow |V| \in \mathcal{O}(|Q| \cdot |\Delta|).
	\end{align*}
	Likewise there is an edge $(v_{(q, q')}, v_{(\tilde{q}, q', a)})$ from $V_1$ to $V_0$ for every
	state $q'$ and transition $(q, a, \tilde{q})$. This are $|Q| \cdot |\Delta|$ edges.
	Together with the edges from $V_0$ to $V_1$, which are limited by $|Q| \cdot |\Delta|$ in the same way,
	it holds that $|E| \in \mathcal{O}(|Q| \cdot |\Delta|)$.
	
	Last $|\{v \in V : p(v) = 1\}| \in \mathcal{O}(|Q|^2)$ is proven. Since vertices with a
	priority of $1$ form a subset of $V_1$, whose size is bounded by $|Q|^2$, also
	$|\{v \in V : p(v) = 1\}| \le |Q|^2$ holds.\\
\end{proof}\quad\\
Next the complexity of the algorithm is received.
\begin{mytheorem}\label{theorem_jurdzcomplexity}
	\algoref{algo_jurdzinski} runs in $\mathcal{O}(|Q|^3 \cdot |\Delta|)$ time
	and $\mathcal{O}(|Q| \cdot |\Delta|)$ space.
\end{mytheorem}
\begin{proof}
	First we proof the space complexity. The algorithm needs to hold the set of successors and
	predecessors of every node $v$, this is limited by the amount of edges $|E|$ of the game graph.
	With \lemmaref{lemma_gamegraphcomplexity} the space complexity follows
	since $|E| \in \mathcal{O}(|Q| \cdot |\Delta|)$.
	
	Next to the time complexity. The initialization in \textbf{lines 1-4} needs time in $\mathcal{O}(|E|)$
	because it processes every outcoming edge, together this are exactly all existing edges.
	The other assignments and the function \textit{succ}$(v)$ itself need to be
	implemented in constant time.
	
	The time needed for processing a vertex $v$ in \textbf{lines 6-22} depends on the number of
	its successors and predecessors. Thus the time complexity is $\mathcal{O}(|pred(v)| + |succ(v)|)$.
	Calculating the \textit{best-nghb-ms}, \textit{nghb-cnt} and the new progress measure of $v$ is
	proportional to the amount of successors. Assuming all \textit{if statements} in \textbf{lines 13-22}
	are implemented in constant time calculating which predecessors needs to be added
	to the working list is proportional to the amount of predecessors obviously.
	Note that the containment test $w \notin L$ in \textbf{line 13} also needs to be implemented
	in constant time. However, this easily can be realized by introducing a
	containment flag for every vertex.
	
	An upper bound gets formed by assuming that every vertex is in the working list. Then,
	the amount of iterations is $|V|$, each consuming time $|pred(v)| + |succ(v)|$.
	Together this visits every edge exactly two times. It follows that the
	time complexity for iterating over each vertex is $\mathcal{O}(|E|)$.
	
	The progress measure of a vertex can increase at most $n + 1$ times
	where $n$ is the amount of vertices with priority $1$. After that its progress measure
	definitely reaches $\infty$. The algorithm only adds a vertex to the working list if it
	will increase its progress measure when processing. It directly follows that each vertex
	can at most be added $n + 1$ times to the working list.
	Using this the upper bound gets extended by assuming that every vertex gets
	added $n + 1$ times to the working list. Iterating over each vertex
	costs $\mathcal{O}(|E|)$ time hence the time complexity for the upper bound
	is $\mathcal{O}(|E| \cdot (n + 1))$. After that each vertex must have reached its
	final progress measure and the program terminates.
	
	Together with the initialization part a time complexity of
	$\mathcal{O}(|E| \cdot (n + 1))$ follows for the whole algorithm.
	
	Last $|E| \in \mathcal{O}(|Q| \cdot |\Delta|)$ and $n \in \mathcal{O}(|Q|^2)$,
	according to \lemmaref{lemma_gamegraphcomplexity}.
	Thus the time complexity is in $\mathcal{O}(|Q|^3 \cdot |\Delta|)$.\\
\end{proof}

% Correctness
\subsubsection{Correctness}\label{section_correctness}
The output of the algorithm is the progress measure function $\mu : V \to \{0, 1, \ldots, n\} \cup \{\infty\}$.
This function is later used to obtain the elements of the simulation relation,
\sectionref{section_compsimrelation} shows this in detail.

Before we talk of correctness of \algoref{algo_jurdzinski}, we define when the
resulting progress measure function $\mu$ is correct.\\\\
Let $M = \{f | f: V \to (\{0, 1, 2\} \cup \{\infty\})\}$ be the set of all functions that map vertices to progress measures.
That are all variants of progress measure functions, note that $\mu$ is a member of the set $M$.
\begin{mydef}
	For all $u \in V$, $\lift_u: M \to M$ defines an unary operator such that
	\begin{align*}
		\lift_u(\mu)(v): = \begin{cases} \mu(v) & \text{if } u \neq v\\
				\text{incr}_{p(v)}(\text{best-nghb-ms}(\mu, v)) & \text{if } u = v\\
			\end{cases}
	\end{align*}
\end{mydef}\quad\\
Given a vertex $u$ the corresponding operator $\lift_u(\mu)$ creates a lifted version of $\mu$.
For all vertices $v \neq u$ the image $\lift_u(\mu)(v)$ is the same as $\mu(v)$,
only the image of $u$ may be changed.
\begin{mydef}
	An element  $x \in X$ is a \textnormal{simultaneous fixed point} of a set of functions $F = \{f | f: X \to X\}$
	iff it is a \textnormal{fixed point} for all those functions. Formally this is $\forall f \in F: f(x) = x$.
\end{mydef}
\begin{mydef}\label{def_correctprogressmeasure}
	A progress measure function $\mu : V \to \{0, 1, \ldots, n\} \cup \{\infty\}$ is a \textnormal{correct result}
	of \algoref{algo_jurdzinski} iff $\mu$ is the \textnormal{least simultaneous fixed point} of all operators $\lift_u$.
\end{mydef}\quad\\
To prove that the result of the algorithm $\mu$ is correct we go more afield. We will use the original
version of \textit{Jurdzi\'nskis algorithm}, from \libref{jurdz_original}, and prove that our,
more efficient, version yields the same result.\\\\
First we define an ordering on the set of progress measure functions.
\begin{mydef}
	Given two functions $\mu_1, \mu_2 \in M$, the \textnormal{progress measure function ordering} $\sqsubseteq$
	defines a partial order over the set $M$. It is given by $\mu_1 \sqsubseteq \mu_2$ iff
	$\mu_1(v) \le \mu_2(v)\,\forall v \in V$.
	
	Analogously $\sqsubset$ defines an order over the set $M$, given by $\mu_1 \sqsubset \mu_2$ iff
	$(\mu_1 \sqsubseteq \mu_2 \land \mu_1 \ne \mu_2)$.
\end{mydef}\quad\\
The progress measure function ordering $\sqsubseteq$ gives the complete lattice structure $\langle M, \sqsubseteq \rangle$.
\begin{myproposition}\label{proposition_liftmonotone}
	The operator $\lift_u$ is $\sqsubseteq$-monotone for all $u \in V$.
\end{myproposition}
\begin{proof}
	Since $\text{incr}_{p(v)}$, for a fixed $p(v)$, is monotonically increasing it holds that
	$\mu \sqsubseteq \lift_u(\mu)$ for all $\mu \in M$.\\
\end{proof}
\begin{mylemma}\label{lemma_liftfixedpoint}
	Every lift operator $\lift_u$ has at least one fixed point.
	\begin{align*}
		\forall u \in V \exists \mu : \lift_u(\mu) = \mu
	\end{align*}
\end{mylemma}
\begin{proof}
	$\langle M, \sqsubseteq \rangle$ forms a complete lattice structure and $\lift_u: M \to M$
	is $\sqsubseteq$-monotone, see \propositionref{proposition_liftmonotone}.
	Using the \textit{Knaster-Tarski theorem} \libref{knaster_tarski} the existance of a
	fixed point follows.\\
\end{proof}\quad\\
The next definition introduces a sequence of progress measure functions. Each received by
applying an arbitrary \textit{lift} operator to the previous element of the sequence.
\begin{mydef}\label{def_sequence}
	Let $\text{fam}: \mathbb{N} \to V$ be a family of elements in $V$ indexed by $\mathbb{N}$.
	$\text{seq}: \mathbb{N} \to M$ is a sequence defined by
	\begin{align*}
		\text{seq}(0)	&= \mu_0\\
		\text{seq}(n + 1)	&= \lift_{\text{fam}(n + 1)}(\text{seq}(n))
	\end{align*}
	where $\mu_0 \in M : \mu_0(v) = 0$ for all $v \in V$.
\end{mydef}\quad\\
Note that a sequence $\text{seq}$ is not necessarily an injective function.
\begin{mycorollary}\label{corollary_seqmonotone}
	\textnormal{seq} is a $\sqsubseteq$-monotone sequence.
\end{mycorollary}
\begin{proof}
	Directly follows from \propositionref{proposition_liftmonotone}.\\
\end{proof}
% Original algorithm
\IncMargin{1em}
\begin{algorithm}
\SetKwData{v}{v}\SetKwData{u}{u}\SetKwData{V}{V}
\SetKwFunction{pm}{$\mu$}\SetKwFunction{lift}{lift}
$\mu: \V \to (\{0, 1, 2\} \cup \{\infty\}), \mu(\v) = 0 \quad \forall \v \in \V$\;
\BlankLine
\BlankLine
\While{$\exists \u \in \V : \pm \sqsubset \lift_{\u}(\pm)$}{
	$\pm := \lift_{\u}(\pm)$\;
}
\BlankLine
\caption{Original version of \textit{Jurdzi\'nskis algorithm} from \libref{jurdz_original}.}\label{algo_jurdzinski_original}
\end{algorithm}\DecMargin{1em}\quad\\
Taking a look on \algoref{algo_jurdzinski_original}, the assignment sequence of $\mu$ in \textbf{line 3}
represents a sequence which we denote by $seq'_{\textit{algo}}$. It is obtained by
$seq'_{\textit{algo}}(i) = \lift_u(seq'_{\textit{algo}}(i - 1))$ for every iteration $i$ of
the \textit{while-loop} in \textbf{line 2-3}.
\begin{mytheorem}\label{theorem_algooriginaltofixpoint}
	After \algoref{algo_jurdzinski_original} has terminated the resulting progress measure function
	$\mu$ is the \textnormal{least simultaneous fixed point} of all operators $\lift_u$.
\end{mytheorem}
\begin{proof}
	First of all $\mu$ is a simultaneous fixed point of all operators $\lift_u$,
	else the algorithm would not have terminated.
	
	$\langle M, \sqsubseteq \rangle$ forms a complete lattice structure and $\lift_u$
	is $\sqsubseteq$-monotone, see \propositionref{proposition_liftmonotone}.
	Using \lemmaref{lemma_liftfixedpoint} and the
	\textit{Knaster-Tarski theorem} \libref{knaster_tarski}, it follows that the
	least simultaneous fixed point of all operators $\lift_u$ does exist.
	If using an approach like \defref{def_sequence} describes,
	it also follows that the first simultaneous fixed point occurring in such a sequence
	is the least simultaneous fixed point.
	
	Since $\mu$ is the first simultaneous fixed point occurring in $seq'_{\textit{algo}}$,
	the proof is complete.\\
\end{proof}\quad\\
Next the connection between \algoref{algo_jurdzinski} and \algoref{algo_jurdzinski_original} is shown.
\begin{myproposition}\label{proposition_algotosequence}
	\algoref{algo_jurdzinski} creates a sequence $\textit{seq}_{\textit{algo}} = \mathbb{N} \to M$.
\end{myproposition}
\begin{proof}
	The progress measure gets initialized with $0$ for all
	vertices in \textbf{lines 1-4}, this corresponds to $\mu_0$ thus
	$\textit{seq}_{\textit{algo}}(0) = \mu_0$.
	In every following iteration, let $\textit{seq}_{\textit{algo}}(n)$ be the current progress measure
	of this round and $v$ the vertex currently processing, \textbf{line 12} corresponds to applying
	$\lift_v(\textit{seq}_{\textit{algo}}(n))$ and assigning it as new progress measure.
	$\textit{seq}_{\textit{algo}}(n + 1)$ is received and this behavior exactly matches
	the definition of a sequence.\\
\end{proof}
\begin{mylemma}\label{lemma_algotofixpoint}
	After applying \algoref{algo_jurdzinski} to the game graph $G^{\star}_{A, A'}$ the resulting
	progress measure $\mu$ is a \textnormal{simultaneous fixed point} of all operators $\lift_v$.
\end{mylemma}
\begin{proof}
	Using \lemmaref{lemma_liftfixedpoint} we follow with
	\textit{Knaster-Tarski theorem} \libref{knaster_tarski} the existance of simultaneous fixed points.
	What is left is the question if the resulting progress measure $\mu$ actually is a simultaneous fixed point.
	In other words, does there exist a $v \in V$ such that $\lift_v(\mu) \sqsubset \mu$?
	The answer is no, we prove by contradiction and let us refer with $\mu'$ to $\lift_v(\mu)$.
	
	Assuming $\exists v \in V : \mu' \sqsubset \mu$.
	Only one value can change by definition, $\mu'(v) \ne \mu(v)$.
	Because $\textit{incr}_{i}(\cdot)$ is monotonically increasing for a fixed $i$, $\mu'(v)$
	must be strict greater than $\mu(v)$. $\mu(v)$ gets assigned in \textbf{line 12} only, this presupposes
	$p(v) \ne 0$. Furthermore this means a successor $s$ of $v$ has increased its
	progress measure in a previous iteration without adding $v$ to the working list.
	\begin{enumerate}
	\item Assuming $v \in V_1$, $v$ would have been added to the
		working list in \textbf{lines 16-17}. \lightning
	\item Assuming $v \in V_0$, $C(v) > 1$ else \textbf{lines 21-22}
		would also add $v$ to the working list. Let $\mu_s$ be the progress measure
		right before $s$ updated its value for the progress measure.
		$C(v) > 1$ thus $\textit{best-nghb-ms}(\mu_s, v) = \textit{best-nghb-ms}(\mu', v)$
		while $\textit{nghb-cnt}(\mu_s, v) > \textit{nghb-cnt}(\mu', v)$ but\\
		$\textit{nghb-cnt}(\mu', v) \ge 1$,
		i.e. $s$ is no optimal choice for $v$ anymore but it has an alternative.
		However, this means $v$ has a successor with a smaller progress
		measure than $s$ and $\textit{best-nghb-ms}$ prefers this successor since $v \in V_0$.
		This contradicts to $\mu'(v) > \mu(v)$ because then $\mu'(v) = \mu(v)$. \lightning
	\end{enumerate}
	It follows that $\lift_v(\mu) = \mu \, \forall v \in V$. So $\mu$ is a simultaneous fixed
	point of all operators $\lift_v$.\\
\end{proof}\quad\\
The following theorem proofs the correctness of \algoref{algo_jurdzinski}.
\begin{mytheorem}\label{theorem_algocorrect}
	The output of \algoref{algo_jurdzinski}, the progress measure function
	$\mu : V \to \{0, 1, \ldots, n\} \cup \{\infty\}$, is a \textnormal{correct result} of the algorithm.
\end{mytheorem}
\begin{proof}
	For $\mu$ to be a correct result it needs to be the least simultaneous fixed point
	of all operators $\lift_u$, compare to \defref{def_correctprogressmeasure}.
	Using \lemmaref{lemma_algotofixpoint}, $\mu$ already is a simultaneous fixed point of
	all operators $\lift_u$.
	
	By using \propositionref{proposition_algotosequence} it is clear that \algoref{algo_jurdzinski}
	uses the same lifting process as the original version uses.
	Furthermore, our algorithm only describes the order of applied
	lifting functions more precise than \algoref{algo_jurdzinski_original}.
	In fact, the original algorithm is a more general version
	of our more efficient implementation and the sequence $\textit{seq}_{\textit{algo}}$
	can also be produced by \algoref{algo_jurdzinski_original}.
	Using \theoremref{theorem_algooriginaltofixpoint} $\mu$ must also be the
	least simultaneous fixed point.
	
	It follows that $\mu$ is a correct result of \algoref{algo_jurdzinski}.\\
\end{proof}

% Simulation from progress measure
\subsection{Simulation from progress measure}\label{section_compsimrelation}
Once the algorithm has finished, the elements of the simulation relation are computed
by using the progress measures.
\begin{mylemma}\label{lemma_algotostrat}
	Let $\mu$ be the progress measure function received by applying \algoref{algo_jurdzinski}
	to the game graph $G^{\star}_{A, A'}$. It holds that there exists a
	\textnormal{winning strategy} for Duplicator in the game $G^{\star}_{A, A'}(q, q')$
	iff $\mu(v_{(q, q')}) < \infty$, where $\star \in \{di, de, f\}$.
\end{mylemma}
\begin{proof}
	\theoremref{theorem_algocorrect} states that the resulting progress measure function
	$\mu$ of the algorithm is a correct result. This implies that it is the least simultaneous fixed point by definition.
	
	\textbf{Theorem 11} from \libref{jurdz_original} describes that the least simultaneous fixed point
	progress measure function $\mu$ forms a \textit{winning set} where $\mu(v_{(q, q')}) < \infty$ holds iff there
	exists a \textit{winning strategy} for Duplicator in $G^{\star}_{A, A'}(q, q')$ with $\star \in \{di, de, f\}$.\\
\end{proof}
\begin{mytheorem}\label{theorem_algotosim}
	After applying \algoref{algo_jurdzinski} to the game graph $G^{\star}_{A, A'}$ the following holds
	\begin{align*}
		q \preceq_{\star} q' \Leftrightarrow \mu(v_{(q, q')}) < \infty \qquad\text{where } \star \in \{di, de, f\}.
	\end{align*}
\end{mytheorem}
\begin{proof}
	Assuming $\mu(v_{(q, q')}) < \infty$, by using \lemmaref{lemma_algotostrat} there
	exists a \textit{winning strategy} for Duplicator in $G^{\star}_{A, A'}$.
	With \lemmaref{lemma_parity_game} the existance of a winning strategy
	implies $q \preceq_{\star} q'$ and vice versa.\\
\end{proof}\quad\\
This connects simulation to the algorithm. The elements of the simulation relation $\preceq_f$ are
efficiently computed by applying \algoref{algo_jurdzinski} to the fair simulation
game graph $G^f_A$ and then using \theoremref{theorem_algotosim}.

% Illustration
\subsection{Illustration}\label{section_illustration}
In this section we illustrate the algorithm and the technique of how a simulation relation is
computed by a progress measure function. Further, we give example automata and explain the process of
creating a game graph and applying \algoref{algo_jurdzinski} to it.\\\\
In an iteration of the algorithm, currently processing with $v$ from the working list,
if the progress measure of $v$ has increased in \textbf{lines 14-22}, the algorithm may reversly build paths
that use $v$ by suggesting it as possible optimal choice for its predecessors.

If it has not increased above the best neighbor measure of predecessor $w$ there is no reason to update
$w$ since it already has a successor that represents a better (or as good as) choice than $v$.
The same applies if the predecessor has priority $0$, it then will reset its counter anyways so there is no
reason why it should especially pick $v$ as successor, any will be optimal.

But whenever predecessor $w \in V_1$, it will pick $v$ as optimal successor if the progress measure of $v$
has increased above the current best neighbor measure of $w$ because $w$ then represents a move by
Spoiler who wants to increase the progress measure. If now $w \in V_0$ and $v$ previously belonged to the best
choices for $w$ its counter of neighbors that have the best progress measure needs to be
decreased. This is because $v$ increased its measure and $w$, which represents a move by Duplicator, wants to build a path that has
a low progress measure. However, if $w$ only had $v$ as best choice $w$ needs to be updated since the increased progress
measure of $v$ will still be the best choice for $w$.\\\\
Everytime we put a predecessor $w$ in the working list it represents that $w$ will choose $v$ as optimal successor that
suits his needs as Duplicator, if $w \in V_0$, or Spoiler, if $w \in V_1$.
If a vertex reaches progress measure $\infty$ it means that his path visits vertices with priority $1$ infinity times while
not visiting priority $0$ that often. Whenever a progress measure reaches $\infty$ it will be passed through all
predecessors to all members on that path since $\forall i \in \{0, 1, \ldots, n\}: i < \infty$.\\\\
For the algorithm we can assume that as soon as we count $n + 1$ occurences of vertices with priority $1$ while not
visiting priority $0$ we have built a path that can visit such vertices infinite times. Because of that we assign $\infty$ as
progress measure whenever increasing a measure that is $n$. Note that the \textit{size of} $\infty$ can be optimized
by analyzing the graph, e.g. by using \textit{SCC} as seen in \sectionref{section_optimization}.

% Examples
\subsubsection{Examples}
To properly understand the algorithm we take a closer look on three examples.\\
% Figure
\begin{figure}[ht]
	 \begin{center}
		\includegraphics[scale=0.5]{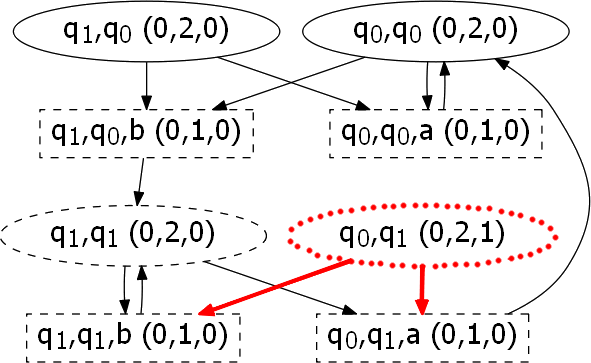}
	\end{center}
	\caption{The game graph from \figref{paritygamegraph}, with similar notation,
		after the algorithm has terminated. The additional tuples indicate the values of the vertex datastructures
		$(\textit{best-nghb-ms}, \textit{nghb-cnt}, \mu)$. The bold edges are the outgoing
		edges of $v_{(q_0, q_1)}$, they indicate the calculation of the \textit{best-nghb-ms}
		of $v_{(q_0, q_1)}$ in the first and only round.}
	\label{algorithmfirstex}
\end{figure}\quad\\
The first is from \figref{paritygamegraph}, in \sectionref{section_gamegraphtogame} we already
saw that Spoiler can not win the game, no matter where the starting position is. Thus and because of \theoremref{theorem_algotosim}
$q_0 \preceq_f q_1$ and $q_1 \preceq_f q_0$. Hence we already know the correct result, after program termination every
vertex must have a progress measure lower than \textit{infinity}.
We also know the amount of vertices with a priority of $1$, it is $n = 1$. This means as soon as a the progress measure $\mu$ of a vertex $v$
that has $\mu(v)= 1$ should be increased it reachs \textit{infinity}.

Looking at \figref{algorithmfirstex} we see the game graph after the algorithm has terminated.
In this example the program will only make one iteration and then terminate. First we notice the default values of a vertex,
for \textit{best-nghb-ms} it is $0$, \textit{nghb-cnt} is the amount of successors and $\mu$ also is $0$.
The vertex $(q_0, q_1)$ initially also had these default values, $(0, 2, 0)$.

The working list of the algorithm is initialized with $(q_0, q_1)$ since this is the only vertex that has a priority of $1$.
In \textbf{lines 6-22} the first iteration starts with $v = (q_0, q_1)$ now. In \textbf{line 10} the \textit{best-nghb-ms}
of $v$ gets calculated based on its succeeding vertices. $v$ has a priority of $1$ and is in $V_1$,
this means it is Spoiler's turn to make a move. Because of that he is looking for the successor
with the highest progress measure, both successors have a progress measure of $0$ so any will be optimal.
Therefore the \textit{best-nghb-ms} of $v$ currently is $0$ and there are two successors having that measure,
\textit{nghb-ms} remains $2$.

\textbf{Line 12} updates the progress measure of $v$ based on its \textit{best-nghb-ms}.
Since the priority of $v$ is $1$ and the \textit{best-nghb-ms} is $0$ the progress measure gets increased from $0$ to $1$.
The program now would inform the predecessors of $v$ about the update but since $v$ has no predecessors the iteration ends.

The working list is empty and the algorithm terminates. $(q_0, q_1)$ managed to increase its progress measure from $0$ to $1$
but not to $\infty$, which would only be one additional update away. $(q_0, q_1)$ has a progress measure below $\infty$
thus $q_1 \preceq_f q_0$ follows, analogously $q_1 \preceq_f q_0$ follows from $(q_1, q_0)$.\\\\
The basic concept is to update the data structures of a vertex based on its successors and then inform
its predecessors about the update, maybe add them to the working list. If, for example, $q \not\preceq_f q'$ then
the algorithm will reversly build the path
\begin{align*}
	\varrho =	&\quad v_0 \rightarrow v_1 \rightarrow \ldots \rightarrow\\
			&\quad \underbrace{v_{(q, q')} \rightarrow \ldots \rightarrow pred(pred(v_{(q, q')})) \rightarrow
				pred(v_{(q, q')}) \rightarrow v_{(q, q')}}_{\text{loop}} \rightarrow \ldots
\end{align*}
by increasing the progress measure of $v_{(q, q')}$, adding a predecessor to the working list which also
adds a predecessor to the working list and so on. This continues and $v_{(q, q')}$ gets visited again through one of
its successors, a loop is created, and it increases its progress measure again.
The process repeats until the progress measure of $v_{(q, q')}$ reaches $\infty$, now all vertices on the loop also
reach $\infty$ and the loop is completely processed.

This creates the path $\varrho$ reversly because we are informing predecessors. $v_0, v_1 \ldots$ may be vertices that
are predecessors of a vertex that is involved in the loop.
The created path has a \textit{lasso}-type structure, first a chain of vertices and then a loop.\\\\
Next the second example which has more iterations.
% Figure
\begin{figure}[ht]
	 \begin{center}
		\includegraphics[scale=0.65]{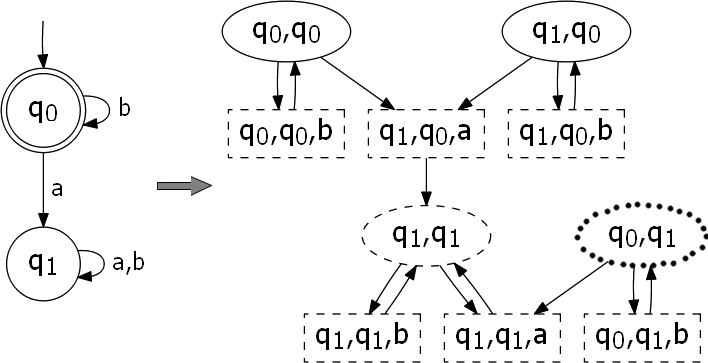}
	\end{center}
	\caption{An automaton $A$ and its game graph $G^f_{A, A}$ with the notation of
		\figref{paritygamegraph}. No states can be merged, we have $q_1 \preceq_f q_0$
		and $q_0 \not\preceq_f q_1$.}
	\label{algorithmsecondex}
\end{figure}
\begin{figure}
	 \begin{center}
		\includegraphics[scale=0.55]{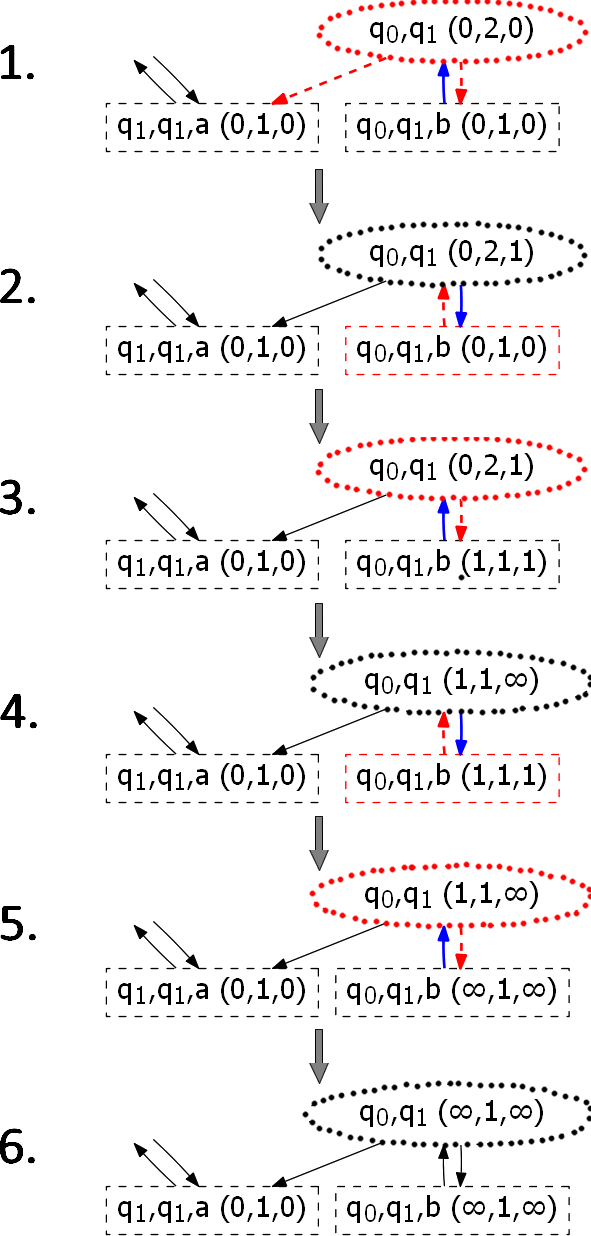}
	\end{center}
	\caption{An excerpt of game graph from \figref{algorithmsecondex}, with similar notation, illustrating
		the six iterations of the algorithm. From the current working vertex bold dashed edges are outgoing and
		bold solid edges are incoming edges indicating the successors of interest for the \textit{best-nghb-ms} and
		the predecessors for update notification.}
	\label{algorithmsecondexsequence}
\end{figure}\quad\\
\figref{algorithmsecondex} shows an automaton with language $\{b^{\omega}\}$.
The language already tells that the two states can not be merged or the new language
will be $\{(ab)^{\omega}\}$. However, $q_1$ is of course redundant as it can not reach a final state.
Note that the presented algorithm is not able to remove $q_1$ as it is not part of a mutually fair simulation.

In \figref{algorithmsecondexsequence} the sequence of iterations the algorithm performs can be seen.
$\infty$ comes after increasing beyond $n = 1$ because there is only one vertex with a priority of $1$.
The working list gets initialized with $v = (q_0, q_1)$, $v$ has two successors and both have
the same progress measure, $0$. Since $v$ has a priority of $1$ its progress measure gets increased to
$1$ and \textbf{lines 13-22} work through the predecessors of $v$. There is only one predecessor, $w = (q_0, q_1, b)$,
and $w$ is in $V_0$ which means this vertex represents a move by Duplicator. Since Duplicator does not want
to increase the progress measure of its vertices he seeks for another possibility to move at instead of $v$. But $w$
has no better alternative, there is no other successor of $w$ with a smaller progress measure, the \textit{nghb-cnt}
of $w$ is $1$. Therefore \textbf{line 21} resolves to true and $w$ gets added to the working list.

In the second iteration the working vertex is $v = (q_0, q_1, b)$. For $v$ the \textit{best-nghb-ms} is the
smallest progress measure of its successors since it is in $V_0$. However, $v$ has only one successor, it has
a progress measure of $1$ so $v$ must increase its progress measure to $2$. We are again in \textbf{lines 13-22} and
work through the predecessors of $v$. There is only $w = (q_0, q_1) \in V_1$ and \textbf{line 16} resolves to true,
$w$ is added to the working list.

This process repeats until $v = (q_0, q_1)$ reaches a progress measure of $\infty$ in the fourth iteration.
It then again forces Duplicator's vertex $w = (q_0, q_1, b)$ to be added into the working list.
After the fifth iteration the cycle ends because \textbf{line 15} resolves to false for $w$.
The reason why is that the progress measure of $v$ has not increased in this round,
the \textit{best-nghb-ms} of $w$ already is $\infty$ from the last iteration thus $\mu(v) \not> B(w)$.

The program terminates because the working list is empty and we obtain two vertices with a progress
measure of $\infty$, $(q_0, q_1)$ and $(q_0, q_1, b)$. We follow $q_0 \not\preceq_f q_1$ but
$q_1 \preceq_f q_0$ since the vertex $(q_1, q_0)$ has a progress measure of $0$.

Note that if $(q_0, q_1)$ would have another predecessor $u$ it would pass its $\infty$, after creating the
loop with $(q_0, q_1, b)$, through $u$. Of course this can only be the case if $u$, which then would represent
a move of Duplicator, has no better alternative to move at.\\\\
The third and last example demonstrates, the possibility of Duplicator to evade an update notification
by choosing a better alternative.
% Figure
\begin{figure}[ht]
	 \begin{center}
		\includegraphics[scale=0.6]{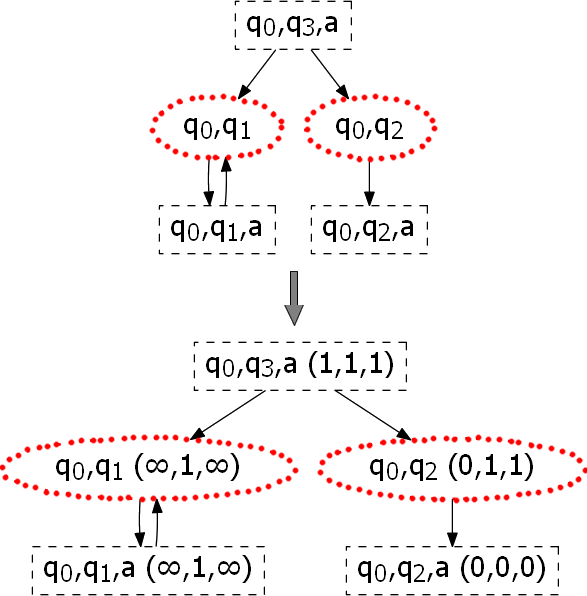}
	\end{center}
	\caption{A theoretical game graph before and after the algorithm was applied
		with the notation of \figref{paritygamegraph}. The figure demonstrates the
		possibility of Duplicator for $(q_0, q_3, a)$ to evade an update
		notification by choosing another edge.}
	\label{algorithmthirdex}
\end{figure}\quad\\
If we apply the algorithm to the game graph from \figref{algorithmthirdex} we initially have a
working list of $\{(q_0, q_1), (q_0, q_2)\}$. We first work with $v = (q_0, q_2)$ which increases its
progress measure to $1$. Next $v$ notifies $w = (q_0, q_3, a)$ about its update but Duplicator has
no interest in increasing the progress measure thus it chooses the vertex $(q_0, q_1)$ as
a better successor than $v$. This event corresponds to \textbf{line 19}, $w$ decreases its \textit{nghb-cnt}
to $1$ since it now only has one neighbor with its current optimal \textit{best-nghb-ms} of $0$ and $w$
gets not added to the working list.

Next its $v = (q_0, q_1)$ turn and he also increases its progress measure to $1$. Its predecessor $(q_0, q_1, a)$ gets
added to the working list. The other predecessor $w = (q_0, q_3, a)$ now also gets added since both of its successors
currently have a progress measure of $1$. For $w$ there is no better alternative to move at as to $v$ or $(q_0, q_2)$.
This represents \textbf{line 21} in the algorithm and the \textit{nghb-cnt} of $w$ was previously decreased
from $2$ to $1$ thus $w$ gets added to the working list.

Now working $(q_0, q_3, a)$ will increase its progress measure to $1$, set its \textit{best-nghb-ms} to $1$ and \textit{nghb-cnt} to $2$.

As next step working $(q_0, q_1, a)$ increases its progress measure and adds $(q_0, q_1)$ again which,
when worked, increases its progress measure to $2$. $v = (q_0, q_1)$ now tries to notify $w = (q_0, q_3, a)$
a second time to pass its progress measure of $2$. Although \textbf{line 18} resolves to true, since $v$ represented one of
the optimal choices to move at in the last round, $w$ gets not added to the working list again because Duplicator better
moves to $(q_0, q_2)$. This is represented by \textbf{line 19} and the \textit{nghb-cnt} of $w$ gets decreased to $1$ again.

The vertex $v = (q_0, q_1)$ together with $(q_0, q_1, a)$ will create a loop and increase their progress measures to $\infty$.
In every iteration $v$ tries to notify $w = (q_0, q_3, a)$ about the update but the \textit{best-nghb-ms} of $w$ is still
only $1$, representing $(q_0, q_2)$ as optimal choice, thus \textbf{line 18} resolves to false (more precisely: $t \ne B(w)$)
in every following iteration.

The program terminates after $v = (q_0, q_1)$ and $(q_0, q_1, a)$ have reached a progress measure of $\infty$ and they
could not pass $\infty$ to $w = (q_0, q_3, a)$. $w$ prefers to move to $(q_0, q_2)$ and accepts its progress
measure of $1$ which is far better than $\infty$ for Duplicator.

% Minimization using fair simulation
\section{Minimization using fair simulation}\label{section_minimization}
In this section we see how to reduce the size of a Büchi automaton using fair simulation.
For this we use two techniques, the first allows us to merge pairs of states and the second technique removes transitions.
First the chain of proof that connects merging and transition removal to fair simulation is constructed.
After that we present the complete minimization algorithm.\\\\
In order to reduce the size of a Büchi automaton $A$ we need to modify it,
merge states or remove transitions. Let $A'$ be the automaton
received after applying the desired modifications. We then evaluate if the language of the automaton did change.
If it does not, $\mathcal{L}(A) = \mathcal{L}(A')$, the modifications are executed and $A := A'$.

Merging of two states $q_1$ and $q_2$ is achieved by adding transitions and then removing one of the states from the automaton.
\sectionref{section_mergestates} describes this process in detail.

Both of our techniques are achieved by adding or removing transitions and
then checking if the language did change.
 
% Language preservation
\subsection{Language preservation}
This section shows how fair simulation is used for merging two states
and the removal of a transition such that it preserves the language.
\begin{mydef}\label{def_inoutpossibilities}
	We define that two states $q$ and $q'$ of a Büchi automaton $A = \langle\Sigma, Q , Q_0, \Delta, F\rangle$
	have the same \textnormal{in- and outgoing possibilities} iff the following holds.
	\begin{align*}
		\begin{pmatrix}
				&\exists a \in \Sigma, q' \in Q : (q_1, a, q') \in \Delta \Rightarrow \exists (q_2, a, q') \in \Delta\\
			\land	&\exists a \in \Sigma, q' \in Q : (q', a, q_1) \in \Delta \Rightarrow \exists (q', a, q_2) \in \Delta\\
			\land	&\exists a \in \Sigma, q' \in Q : (q_2, a, q') \in \Delta \Rightarrow \exists (q_1, a, q') \in \Delta\\
			\land	&\exists a \in \Sigma, q' \in Q : (q', a, q_2) \in \Delta \Rightarrow \exists (q', a, q_1) \in \Delta
		\end{pmatrix}
	\end{align*}
	Next we define the relation $\sim$ between states to hold iff both states have the
	same \textnormal{in- and outgoing possibilities}.
\end{mydef}
\begin{mylemma}\label{lemma_safemerge}
	Given two states $q_1, q_2 \in Q$ of a Büchi automaton $A =\\
	\langle\Sigma, Q , Q_0, \Delta, F\rangle$ where
	$A'$ is the automaton obtained by removing $q_2$ (or $q_1$ if $q_2 \in F \land q_1 \notin F$) and all
	its in- and outgoing transitions, i.e. merging states $q_1$ and $q_2$. the following holds.
	\begin{align*}
		q_1 \sim q_2 \Rightarrow \mathcal{L}(A) = \mathcal{L}(A')
	\end{align*}
	I.e. merging $q_1$ and $q_2$ does not change the language if they have the same in- and outgoing possibilities.
\end{mylemma}
\begin{proof}
	We proof by contradiction, assuming $\mathcal{L}(A) \neq \mathcal{L}(A')$.
	\begin{itemize}
	\item[1.]
		$\mathcal{L}(A) \supset \mathcal{L}(A') \Rightarrow \exists w=a_1a_2a_3\ldots :
		w \in \mathcal{L}(A) \land w \notin \mathcal{L}(A')$\\
		Let $\pi = q'_0a_1q'_1a_2q'_2a_3q'_3\ldots$ be an arbitrary run that corresponds to $w$ and
		$w.l.o.g.$ the states where merged by removing $q_2$. We construct $\pi'$ from $\pi$
		by exchanging $q_2$ with $q_1$.
		$\pi'$ is still a valid run since if $\pi$ uses $(q', a, q_2)$ or $(q_2, a, q')$ it follows that the
		transition $(q', a, q_1)$ or $(q_1, a, q')$ does also exist by the left hand side of the lemma.
		But $q_1$ and its transitions do also exist in $A'$ because we only removed $q_2$.
		We follow that $\pi'$ corresponds to $w$, forms an accepting run and $w \in \mathcal{L}(A')$. \lightning
	\item[2.]
		$\mathcal{L}(A) \subset \mathcal{L}(A') \Rightarrow \exists w=a_1a_2a_3\ldots :
		w \notin \mathcal{L}(A) \land w \in \mathcal{L}(A')$\\
		Let $\pi' = q'_0a_1q'_1a_2q'_2a_3q'_3\ldots$ be an arbitrary run that corresponds to $w$ and
		$w.l.o.g.$ the states where merged by removing $q_2$. We immediately see that $\pi = \pi'$ forms an
		accepting run in $A$ because every existing transition or state from $A'$ also exists in $A$.
		We follow that $w \in \mathcal{L}(A)$. \lightning
	\end{itemize}
	This follows $\mathcal{L}(A) = \mathcal{L}(A')$, the language did not change.\\
\end{proof}\quad\\
Using \lemmaref{lemma_safemerge} we can safely merge two states if both have the same in- and outgoing
possibilities (\defref{def_inoutpossibilities}).
But what if this is not the case? For example if there is a transition $(q_1, a, q_3)$ and the corresponding transition
$(q_2, a, q_3)$ does not exist.
We need to add $(q_2, a, q_3)$ to the automaton and check wether this changed the language.
This is where the correlation to simulation shows.\\\\
So far fair simulation was only used for states from one automaton. If two automata $A$ and $A'$ only
differ in their set of transitions we now use fair simulation between two automata. For example
$q \preceq_f q'$ where $q$ is a state from $A$ and $q'$ a state from $A'$.
Recalling \defref{def_simulation}, fair simulation is easily extended by letting the left hand side of the implications
use transitions of $A$ and the right hand side transitions from $A'$.\\\\
Using this we define fair simulation between automata.
\begin{mydef}\label{def_automatasimulation}
	Given two Büchi automata $A = \langle\Sigma, Q, Q_0, \Delta, F\rangle$ and
	$A' = \langle\Sigma, Q, Q'_0, \Delta', F\rangle$ where $Q_0 = Q'_0$ we say
	\begin{align*}
		 A' \textnormal{ fairly simulates } A \text{ iff } \forall q \in Q_0 \exists q' \in Q'_0 : q \preceq_f q'
	\end{align*}
\end{mydef}\quad\\
Next we see how fair simulation between automata preserves language.
\begin{mycorollary}\label{corollary_smallinclusion}
	Given a Büchi automaton $A = \langle\Sigma, Q, Q_0, \Delta, F\rangle$ and two states $q, q' \in Q$ it holds that
	$q \preceq_f q' \Rightarrow \mathcal{L}(A^q) \subseteq \mathcal{L}(A^{q'})$ 
\end{mycorollary}
\begin{proof}
	Follows directly from the definition of fair simulation as described in \defref{def_simulation}.\\
\end{proof}
\begin{mylemma}\label{lemma_biginclusion}
	Given two Büchi automata $A = \langle\Sigma, Q , Q_0, \Delta, F\rangle$, $A' =\\
	\langle\Sigma, Q, Q_0', \Delta', F\rangle$ where $Q_0 = Q_0'$, the following holds
	\begin{align*}
		A' \text{ fairly simulates } A \Rightarrow \mathcal{L}(A) \subseteq \mathcal{L}(A')
	\end{align*}
\end{mylemma}
\begin{proof}
	Since $A'$ fairly simulates $A$ we know that there exists a $q' \in Q'_0$ for each $q \in Q_0$ so that $q \preceq_f q'$.
	With \corollaryref{corollary_smallinclusion} $\mathcal{L}(A^q) \subseteq \mathcal{L}(A'^{q'})$ follows.
	\begin{align*}
		\mathcal{L}(A) = \bigcup\limits_{q \in Q_0} \mathcal{L}(A^{q})
		\subseteq \bigcup\limits_{q' \in Q'_0} \mathcal{L}(A'^{q'}) = \mathcal{L}(A').
	\end{align*}
\end{proof}
\begin{mytheorem}\label{theorem_languageequivalence}
	$A = \langle\Sigma, Q , Q_0, \Delta, F\rangle$, $A' = \langle\Sigma, Q, Q_0', \Delta', F\rangle$
	are two Büchi automata and $Q_0 = Q_0'$. Then, if both automata \textnormal{fairly simulate}
	each other, their language is the same.
	\begin{align*}
		(A' \text{ fairly simulates } A \land A \text{ fairly simulates } A') \Rightarrow \mathcal{L}(A) = \mathcal{L}(A')
	\end{align*}
\end{mytheorem}
\begin{proof}
	$\mathcal{L}(A) \subseteq \mathcal{L}(A') \land \mathcal{L}(A) \supseteq \mathcal{L}(A') \Rightarrow \mathcal{L}(A) = \mathcal{L}(A')$.
	Using \lemmaref{lemma_biginclusion} we are done.\\
\end{proof}\quad\\
For language equality we need inclusion in both directions, $\mathcal{L}(A) \subseteq \mathcal{L}(A')$
and $\mathcal{L}(A) \supseteq \mathcal{L}(A')$. So if we add transitions for an attempted merge
we need to check if $A$ fairly simulates $A'$ and if also $A'$ fairly simulates $A$. The same applies if we want to
remove transitions, language does not change if we have fair simulation in both directions between
the automaton before and after the modification.

% Modifing the game graph
\subsection{Modifing the game graph}
For checking language inclusion we use a method that allows us to efficiently modify the
game graph.

We present tools for game graph manipulation and in the next section take a
closer look on how they are used to compute language inclusion.

\begin{mydef}
	For a given Büchi automaton $A = \langle\Sigma, Q , Q_0, \Delta, F\rangle$, a game graph
	$G^f_{A, A'} = \langle V_0, V_1, E, p\rangle$ and a set of transitions
	$T \subseteq Q \times \Sigma \times Q$, we make the following four definitions.
	\begin{enumerate}
		\item $\textnormal{rem}(A, T) = \langle\Sigma, Q , Q_0, \Delta \setminus T, F\rangle$ is the
			automaton without transitions in $T$.
		\item $\textnormal{rem}(G^f_{A, A'}, T)$ is the game graph $\langle V_0, V_1, E', p\rangle$ where
			\subitem $E' = E \setminus \{\underbrace{\left(v_{(q, q', a)}, v_{(q, \tilde{q})}\right)}_{\text{transition of Duplicator}}
				| v_{(q, q', a)} \in V_0, v_{(q, \tilde{q})} \in V_1,$\\
				\subsubitem $(q', a, \tilde{q}) \in T\}$.
		\item $\textnormal{add}(A, T) = \langle\Sigma, Q , Q_0, \Delta \cup T, F\rangle$ is the automaton $A$
			additionally with transitions in $T$.
		\item $\textnormal{add}(G^f_{A, A'}, T)$ is the game graph $\langle V_0, V_1, E', p\rangle$ where
			\subitem $E' = E \cup \{\underbrace{\left(v_{(q, q')}, v_{(\tilde{q}, q', a)}\right)}_{\text{transition for Spoiler}}
				| v_{(q, q')} \in V_1,$\\
				\subsubitem $v_{(\tilde{q}, q', a)} \in V_0, (q, a, \tilde{q}) \in T\}$.
	\end{enumerate}
\end{mydef}\quad\\
As we see \textit{rem} removes transitions, given as set $T$, only from Duplicator's automaton
and \textit{add} adds transitions only to Spoiler's automaton.
\begin{mylemma}\label{lemma_modfiedgraph}
	Given two given Büchi automaton $A = \langle\Sigma, Q , Q_0, \Delta, F\rangle$ and $A' = \langle\Sigma, Q , Q_0, \Delta', F\rangle$,
	where $T$ is a set of transitions.
	If $\Delta' = \Delta \setminus T$ then $G^f_{A, A'} = \textnormal{rem}(G^f_{A, A}, T)$.
	If $\Delta' = \Delta \cup T$ then  $G^f_{A', A} = \textnormal{add}(G^f_{A, A}, T)$.
\end{mylemma}
\begin{proof}
	Since $\text{add}(A, T) = A'$ we directly follow that the game graph constructed by $G^f_{A', A}$ is the same as that generated
	by modifying $G^f_{A, A}$ by using $\text{add}(G^f_{A, A}, T)$. This is because \textit{add} only adds transitions to Spoiler,
	the same transitions that would be generated by giving Spoiler the automaton $A'$ for use from beginning.
	
	Analogue for \textit{rem} where Duplicator looses the same transitions that would be left if he directly started with
	the modified automaton $A'$.\\
\end{proof}
\begin{mycorollary}
	Furthermore, given two Büchi automata $A$, $A'$ and two sets of transitions $T$, $T'$, we have
	\begin{enumerate}
		\item $G^f_{A, \textnormal{rem}(A, T)} = \textnormal{rem}(G^f_{A, A}, T)$ and
			\subitem $\textnormal{rem}(\textnormal{rem}(G^f_{A, A}, T), T') = \textnormal{rem}(G^f_{A, A}, T \cup T')$.
		\item $G^f_{\textnormal{add}(A, T), A} = \textnormal{add}(G^f_{A, A}, T)$ and
			\subitem $\textnormal{add}(\textnormal{add}(G^f_{A, A}, T), T') = \textnormal{add}(G^f_{A, A}, T \cup T')$.
	\end{enumerate}
\end{mycorollary}
\begin{proof}
	Directly follows from \lemmaref{lemma_modfiedgraph}.\\
\end{proof}

% Merge states
\subsection{Merge states}\label{section_mergestates}
We have seen that a pair of states $(q, q')$ can be merged without changing the language
of the automaton when they directly or delayedly simulate each other. For fair simulation this is not the case.
However, chances are high in practice that such a pair can also be merged without changing the language.
We say $(q, q')$ is \textit{of interest for merging} if the states fairly simulate each other,
\begin{align*}
	q \preceq_f q' \land q' \preceq_f q.
\end{align*}
Given such a pair we construct the automaton $A'$ where both states have the same
in- and outgoing possibilites, i.e. $q \sim q'$ (\defref{def_inoutpossibilities}).

In detail we define $A'$ as the automaton $add(A, T)$ where
\begin{align*}
	T = \{(q, a, \tilde{q}) \notin \Delta | (q', a, \tilde{q}) \in \Delta\}\\
		\cup \{(\tilde{q}, a, q) \notin \Delta | (\tilde{q}, a, q') \in \Delta\}\\
		\cup \{(q', a, \tilde{q}) \notin \Delta | (q, a, \tilde{q}) \in \Delta\}\\
		\cup \{(\tilde{q}, a, q') \notin \Delta | (\tilde{q}, a, q) \in \Delta\}	&.
\end{align*}
Recalling \theoremref{theorem_languageequivalence} we need to compute if $A'$ fairly simulates $A$ and if $A$ fairly simulates $A'$.\\\\
The first computation is easy. We already know that $A'$ fairly simulates $A$ because $A'$ has more transitions than $A$.
Hence it intuitively has more possibilities.

For $A'$ to fairly simulate $A$ we need to show $\forall q \in Q_0 \exists q' \in Q'_0 : q \preceq_f q'$. We simply select $q' = q$
and $q \preceq_f q'$ holds because if $q$ in $A$ builds an accepting run $\pi$ then $q'$ in $A'$ can build the same run
$\pi' = \pi$. All transitions from $A$ are also available for $q'$ in $A'$.\\\\
The second computation is not trivial, to compute if $A$ fairly simulates $A'$ we construct the game graph $G^f_{A', A} = add(G^f_A, T)$.
This is the game graph where Spoiler plays on $A'$ and Duplicator on $A$.
\begin{mytheorem}\label{theorem_modificatedsimulation}
	Let $\mu$ be the progress measure function obtained by applying \algoref{algo_jurdzinski} to the graph
	without modifications $G^f_A$.
	
	Also let $\mu'$ be the progress measure function obtained by applying the algorithm to the modified game graph
	$\text{add}(G^f_A, T) = G^f_{A', A}$, where $T$ is a set of transitions.
	Then the following holds.
	\begin{align*}
		A \text{ fairly simulates } A' \text{ iff } \forall v \in V : (\mu(v) \neq \infty \Rightarrow \mu'(v) \neq \infty)\\
		\land (\mu(v) = \infty \Rightarrow \mu'(v) = \infty)
	\end{align*}
\end{mytheorem}
\begin{proof}
	If the right hand side holds we directly follow
	\begin{align*}
		\forall q' \in Q'_0 \exists q \in Q_0 : q' \preceq_f q
		\overset{\defref{def_automatasimulation}}{\Longrightarrow} A \text{ fairly simulates } A'
	\end{align*}
	since we trivially have $q_i \preceq_f q_i \forall i$ in $G^f_A$ and also in $G^f_{A', A}$, we simply select $q = q'$.
	
	If $A$ fairly simulates $A'$ we have $\forall q' \in Q'_0 \exists q \in Q_0 : q' \preceq_f q$,
	i.e. the trivial pairs $q'_i \preceq_f q_i$ where $q'_i = q_i$.
	Since $\preceq_f$ transitive, which easily can be seen, we build all other
	elements $\tilde{q} \preceq_f q_i$ of the simulation relation with $\tilde{q} \preceq_f q'_i$ by using
	\begin{align*}
		\tilde{q} \preceq_f q'_i \land q'_i \preceq_f q_i \Rightarrow \tilde{q} \preceq_f q_i.
	\end{align*}
\end{proof}\quad\\
We see that if we apply \algoref{algo_jurdzinski} to $G^f_{A', A}$ and the obtained simulation relation
is the same as after running the algorithm on $G^f_A$, $A$ fairly simulates $A'$.
Together with the trivial statement that $A'$ fairly simulates $A$ we receive language equivalence
$\mathcal{L}(A) = \mathcal{L}(A')$ by \theoremref{theorem_languageequivalence}.
Both states then have the same in- and outgoing possibilities, we
apply \lemmaref{lemma_safemerge} and merge the two states without changing the language.

Summarized we first search for pairs $(q, q')$ where $q \preceq_f q'$ and $q' \preceq_f q$.
Then we construct $G^f_{A', A}$ and run \algoref{algo_jurdzinski} on it.
We compare results and if both simulation relations are equal the merge is accepted
and does not change the language of the automaton.

% Remove redundant transitions
\subsection{Remove redundant transitions}\label{section_removetransitions}
A transition $(q, a, q')$ is of interest for removal if
\begin{align*}
	\exists \tilde{q} \in Q : (q, a, \tilde{q}) \in \Delta \land q' \preceq_f \tilde{q}.
\end{align*}
In practice chances are high that such transitions do not change the language when
removed, compared to arbitrary transitions. Given such a transition we construct the
automaton $A'$ where this transition was removed. In detail $A' = rem(G^f_A, T)$ where $T = \{(q, a, q')\}$.

We again recall \theoremref{theorem_languageequivalence} and compute if $A'$ fairly simulates $A$ and $A$ fairly simulates $A'$.\\\\
This time the second computation is easy. If we remove transitions from $A$ we already know that $A$ fairly
simulates the received automaton $A'$ because $A'$ intuitively has less possibilities.

For $A$ to fairly simulate $A'$ we need to show $\forall q' \in Q'_0 \exists q \in Q_0 : q' \preceq_f q$.
We select $q = q'$ and $q' \preceq_f q$ holds because if $q'$ in $A'$ builds an accepting run $\pi'$
then $q$ in $A$ can build the same path $\pi = \pi'$. All transitions from $A'$ are also available for
$q$ in $A$, i.e. for all transitions $(q', a, \tilde{q})$ in $A'$ there does also exist the transition
$(q, a, \tilde{q})$ in $A$, analogously for transitions $(\tilde{q}, a, q')$ in $A'$ there exists the
transition $(\tilde{q}, a, q)$ in $A$.\\\\
The first computation is not trivial, we proceed likewise to \sectionref{section_mergestates}.
In order to compute if $A'$ fairly simulates $A$, we consruct the game graph $G^f_{A, A'} = rem(G^f_A, T)$ .
In this game graph Spoiler plays on $A$ and Duplicator on the modified automaton $A'$.
\begin{mytheorem}\label{theorem_modificatedsimulationremove}
	Let $\mu$ be the progress measure function obtained by applying \algoref{algo_jurdzinski}
	to the graph without modifications $G^f_A$.
	
	Also let $\mu'$ be the progress measure function obtained by applying the algorithm to the
	modified game graph $\text{rem}(G^f_A, T) = G^f_{A, A'}$, where $T$ is a set of transitions.
	Then the following holds.
	\begin{align*}
		A' \text{ fairly simulates } A \text{ iff } \forall v \in V : (\mu'(v) \neq \infty \Rightarrow \mu(v) \neq \infty)\\
		\land (\mu'(v) = \infty \Rightarrow \mu(v) = \infty)
	\end{align*}
\end{mytheorem}
\begin{proof}
	Analogously as for \theoremref{theorem_modificatedsimulation}.\\
\end{proof}\quad\\
If both simulation relations are the same we receive that $A'$ fairly simulates $A$,
together with $A$ fairly simulates $A'$ we now obtain language equivalence, $\mathcal{L}(A) = \mathcal{L}(A')$
for removing the transition.

Summarized we first search for transitions of interest $(q, a, q')$. Then we construct $G^f_{A, A'}$
and run \algoref{algo_jurdzinski} on it. We compare results and if both simulation relations are equal the
transition removal is accepted and does not change the language of the automaton.

% Algorithm
\subsection{Algorithm}
Take a look at \algoref{algo_minimization}, first presented in \libref{fair_minimization},
the complete minimization algorithm that uses the introduced techniques to reduce the size
of a Büchi automaton using fair simulation.\\\\
% Complete algorithm implementation
\IncMargin{1em}
\begin{algorithm}
\SetKwData{v}{v}\SetKwData{V}{V}\SetKwData{L}{L}\SetKwData{T}{T}
\SetKwData{S}{S}\SetKwData{A}{A}
\SetKwFunction{pm}{$\mu$}\SetKwFunction{add}{add}\SetKwFunction{rem}{rem}
\SetKwFunction{merge}{merge}\SetKwFunction{remove}{remove}
$G^f_{\A} := \langle V_0, V_1, E, p \rangle$\;
$\V := V_0 \cup V_1$\;
$\pm := \algoref{algo_jurdzinski}(G^f_{\A})$\;
\BlankLine
\BlankLine
$\L_1 := \emptyset$\;
\For{$\v_{(q, q')} \in \V : \mu(\v_{(q, q')}) < \infty $}{
	\If{$\pm(\v_{(q', q)}) < \infty \land \v_{(q', q)} \notin \L_1$}{
		$\L_1 := \L_1 \cup \v_{(q, q')}$\;
	}
}
\BlankLine
$\L_2 := \emptyset$\;
\For{$(q, a, q') \in \Delta : \left(\exists \tilde{q} \in Q : (q, a, \tilde{q}) \in \Delta \land \pm(\v_{(q', \tilde{q})}) < \infty\right)$}{
	$\L_2 := \L_2 \cup (q, a, q')$\;
}
\BlankLine
\BlankLine
$\S_1 := \emptyset$\;
\For{$\v_{(q, q')} \in \L_1$}{
	$\T :=\{(q, a, \tilde{q}) \notin \Delta | (q', a, \tilde{q}) \in \Delta\}$\;
	\qquad$\cup \{(\tilde{q}, a, q) \notin \Delta | (\tilde{q}, a, q') \in \Delta\}$\;
	\qquad$\cup \{(q', a, \tilde{q}) \notin \Delta | (q, a, \tilde{q}) \in \Delta\}$\;
	\qquad$\cup \{(\tilde{q}, a, q') \notin \Delta | (\tilde{q}, a, q) \in \Delta\}$\;
	$\pm' := \algoref{algo_jurdzinski}(\add(G^f_{\A}, \T))$\;
	\If{$\forall \v \in \V : \left(\pm'(\v) \neq \infty \Rightarrow \pm(\v) \neq \infty\right) \newline\hphantom\qquad\qquad
		\land \left(\pm'(\v) = \infty \Rightarrow \pm(\v) = \infty\right)$}{
		$\S_1 := \S_1 \cup \v_{(q, q')}$\;
	}
}
\BlankLine
$\S_2 := \emptyset$\;
\For{$(q, a, q') \in \L_2$}{
	$\T := \{(q, a, q')\}$\;
	$\pm' := \algoref{algo_jurdzinski}(\rem(G^f_{\A}, \T))$\;
	\If{$\forall \v \in \V : \left(\pm'(\v) \neq \infty \Rightarrow \pm(\v) \neq \infty\right) \newline\hphantom\qquad\qquad
		\land \left(\pm'(\v) = \infty \Rightarrow \pm(\v) = \infty\right)$}{
		$\S_2 := \S_2 \cup (q, a, q')$\;
	}
}
\BlankLine
\BlankLine
\For{$\v_{(q, q')} \in \S_1$}{
	\A := $\merge(q, q', \A)$\;
}
\BlankLine
\For{$(q, a, q') \in \S_2$}{
	 \A:= $\remove((q, a, q'), \A)$\;
}
\BlankLine
\Return A\;
\BlankLine
\caption{Complete algorithm for minimization of Büchi automaton
	$A = \langle \Sigma, Q, Q_0, \Delta, F \rangle$ using fair simulation.}\label{algo_minimization}
\end{algorithm}\DecMargin{1em}\quad\\
\textbf{Lines 1-3} construct the game graph and apply the initial run of \algoref{algo_jurdzinski}.

In \textbf{lines 4-10} we compute the states of interest for merging and store it in $L_1$.
We store the transitions of interest for removal in $L_2$. As specified in
\sectionref{section_mergestates} and \sectionref{section_removetransitions}.

\textbf{Lines 11-25} check if merging or transition removal would change the language of the automaton.
If a merge would not change the language the pair gets stored in $S_1$, and $S_2$ stores the
transitions that would not change the language if removed.
In \textbf{lines 13-16} the set $T$ describes the transitions that need to be added for
$q$ and $q'$ to have the same in- and outgoing possibilities as seen in \sectionref{section_mergestates}.
\textbf{Line 18 and 24} represent \theoremref{theorem_modificatedsimulation} and
\theoremref{theorem_modificatedsimulationremove} respectively.

\textbf{Lines 26-29} merge pairs of states and remove transitions that do not change the
language if merged or removed.

% Complexity
\subsubsection{Complexity}
Before we start to analyze the complexity of the minimization algorithm let us establish some assumptions.
We assume that checking the conditions in \textbf{lines 5, 6, 9, 18, 24} can be done in constant time.
Most of them can be checked while proccessing \algoref{algo_jurdzinski} and, for example, $L_1$ can be
implemented as \textit{Hashset} which allows membership test in constant time.

Also we assume that we only hold one game graph in the memory. If we need to change the game graph,
in \textbf{lines 17, 23}, we do not create a new game graph and instead modify the original graph.
If the modification changes the language we undo the changes, this is described in
\sectionref{section_optimization} in more detail.
\begin{mytheorem}\label{theorem_minimizationcomplexity}
	Given a Büchi automaton $A = \langle \Sigma, Q, Q_0, \Delta, F\rangle$
	\algoref{algo_minimization} runs in $\mathcal{O}(|Q|^4 \cdot |\Delta|^2)$ time
	and $\mathcal{O}(|Q| \cdot |\Delta|)$ space.
\end{mytheorem}
\begin{proof}
	Let us first analyze the space complexity. \theoremref{theorem_jurdzcomplexity} states
	that \algoref{algo_jurdzinski} runs in $\mathcal{O}(|Q| \cdot |\Delta|)$ space.
	But we first need to construct the game graph and hold it in memory.
	However, the game graph also has a size of $|Q| \cdot |\Delta|$ as we know
	from \lemmaref{lemma_gamegraphcomplexity}. It will not increase the space complexity.
	
	Lists $L_1, L_2, S_1, S_2$ store vertices and transitions which are of interest.
	Their size depends on the size of the game graph, $|Q| \cdot |\Delta|$.
	The progress measure function $\mu$ needs to be holded in memory the whole time.
	Its domain is $V$, the vertices of the game graph, but $|V| \in \mathcal{O}(|Q| \cdot |\Delta|)$.
	So in total we have a space complexity of $\mathcal{O}(|Q| \cdot |\Delta|)$.

	For the time complexity we first spot \algoref{algo_jurdzinski} runs in
	$\mathcal{O}(|Q|^3 \cdot |\Delta|)$ time as \theoremref{theorem_jurdzcomplexity} states.
	
	For constructing the initial game graph $G^f_A$ we need to iterate over each state and each
	transition especially to create vertices $v_{(q, q', a)} \in V_0$. Therefore we have a time complexity
	of $\mathcal{O}(|Q| \cdot |\Delta|)$ for \textbf{line 1}.
	
	Last we need to know how often we apply \algoref{algo_jurdzinski} in \textbf{lines 17, 23},
	let us refer with $k$ to this amount. Obviously $k$ is proportional to the number of attempted
	merges and transition removals. Those are dependent on the amount of simulation
	relation elements $\preceq_f$ and transitions $\Delta$.
	Since $\preceq_f : Q \times Q$ we follow $k \in \mathcal{O}(|Q|^2 + |\Delta|) \subseteq
	\mathcal{O}(2 \cdot |Q| \cdot |\Delta|)$ because $|Q|^2 \le |Q| \cdot |\Delta| \land
	\Delta \le |Q| \cdot |\Delta|$ as we know by \lemmaref{lemma_gamegraphcomplexity}.
	
	In total we have a time complexity of $\mathcal{O}(|Q|^3 \cdot |\Delta| \cdot k) \subseteq
	\mathcal{O}(|Q|^4 \cdot |\Delta|^2)$.\\
\end{proof}

% Examples
\subsection{Examples}
Let us consider two examples. The first example is the automaton of \figref{fair_fail}.
We see its game graph after the initial run of \algoref{algo_jurdzinski} in \figref{algorithmfirstex}.
The algorithm yields $q_0 \preceq_f q_1$ and $q_1 \preceq_f q_0$ so this pair is
of interest for merging.

Next we need to compute $T$, the set of transitions that need to be added for $q_0$ and $q_1$
to have the same in- and outgoing possibilities. $q_0$ has predecessors $q_0$ and $q_1$ both with
letter $a$. $q_1$ now also needs this predecessors, we add $(q_0, a, q_1)$ and $(q_1, a, q_1)$ to $T$.
$q_0$ has successors $q_0$, with $a$, and $q_1$, with $b$. $q_1$ already has both required transitions.
Now we move to $q_1$, it has two predecessors $q_0$ and $q_1$ both with letter $b$, we need to add
$(q_0, b, q_0)$ and $(q_1, b, q_0)$ to $T$. Since $q_0$ already has all transitions required by the
successors of $q_1$ we are done and
\begin{align*}
	T = \{(q_0, a, q_1), (q_1, a, q_1), (q_0, b, q_0), (q_1, b, q_0)\}.
\end{align*}
% Figure
\begin{figure}[ht]
	 \begin{center}
		\includegraphics[scale=0.6]{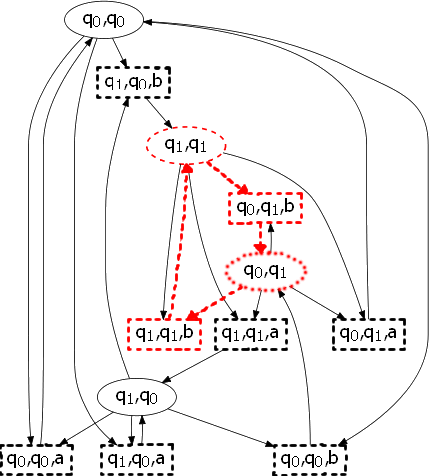}
	\end{center}
	\caption{The game graph $add(G^f_A, T)$ where $A$ is from \figref{fair_fail} and $T$ the
		transitions left for $q_0$ and $q_1$ to have the same in- and outgoing possibilities
		with the notation of \figref{paritygamegraph}. The dashed edges indicate the loop,
		starting at $(q_0, q_1)$ which increases the progress
		measure for every vertex to $\infty$.}
	\label{fullalgorithmfirstex}
\end{figure}\quad\\
Using $add(G^f_A, T)$ creates the game graph seen in \figref{fullalgorithmfirstex}.
When running \algoref{algo_jurdzinski} on the modified game graph every vertex reaches
a progress measure of $\infty$. The resulting simulation relation $\preceq_f$ has no elements,
there is no fair simulation anymore. By \theoremref{theorem_modificatedsimulation} we know this
means the language changes if we merge $q_0$ and $q_1$.

\algoref{algo_minimization} does not accept the attempted merge. Since there are no
more merge candidates and no candidates for transition removal the program terminates.\\\\
In the next example we see the successful removal of a transition.
% Figure
\begin{figure}[ht]
	 \begin{center}
		\includegraphics[scale=0.5]{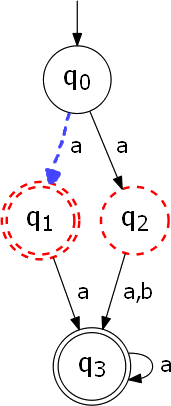}
	\end{center}
	\caption{An automaton where $q_1 \preceq_f q_2$ , if \algoref{algo_minimization} is applied,
	leads to the removal of transition $(q_0, a, q_1)$ because $(q_0, a, q_2)$ exists.
	The attempted removal does not change the language and gets accepted.}
	\label{fullalgorithmsecondex}
\end{figure}\quad\\
Looking at \figref{fullalgorithmsecondex} we have an automaton with $q_1 \preceq_f q_2$,
$q_3 \preceq_f q_0$ and $q_3 \preceq_f q_2$ which clearly can be seen.
Although there is no pair of states for merging we can remove a transition.

In \textbf{line 9} of \algoref{algo_minimization} we find the transition $(q_0, a, q_1)$ since
there does exist transition $(q_0, a, q_2)$ and $q_1 \preceq_f q_2$.
When building the game graph $rem(G^f_A, \{(q_0, a, q_1)\})$ in \textbf{line 23} we
remove the edges
\begin{align*}
	(v_{(q_1, q_0, a)}, v_{(q_1, q_1)}), (v_{(q_2, q_0, a)}, v_{(q_2, q_1)})
	\text{ and } (v_{(q_3, q_0, a)}, v_{(q_3, q_1)})
\end{align*}
from the original graph $G^f_A$.
As we see those edges are exactly the edges of Duplicator that would use
the removed transition $(q_0, a, q_1)$.

If we now apply \algoref{algo_jurdzinski} on the modified game graph the resulting simulation
relation is the same as before. By \theoremref{theorem_modificatedsimulationremove}
this means the language does not change if $(q_0, a, q_1)$ is removed from the automaton.
The attempted removal is accepted and the program terminates.

\newpage

% Optimization
\section{Optimization}\label{section_optimization}
This section presents some optimizations for the algorithm.

% Reuse
\subsubsection*{Reuse}
When using \algoref{algo_jurdzinski} in \textbf{line 3} of \algoref{algo_minimization} in order to
compute the simulation relation, we can reuse information for future runs of the algorithm in \textbf{lines 17, 23}.
In practice modifcations on the game graph are small and most progress measures of vertices will not change.

Instead of initializing the progress measure of vertices or the arrays $B$ and $C$ with the default assignment,
we can reuse information gathered from the first run of \algoref{algo_jurdzinski} in \textbf{line 3}.
The following lemma examines this more closely.
\begin{mylemma}\label{lemma_pmincreases}
	For the progress measures functions $\mu$ and $\mu'$ received from $G^f_A$ and $\text{add}(G^f_A, T)$,
	where $T$ is a set of transitions, the following holds.
	\begin{align*}
		\forall v \in V : \mu(v) \le \mu'(v)
	\end{align*}
	The same applies to $\text{rem}(G^f_A, T)$.
	I.e. the progress measure can only get greater after modifying the graph.
\end{mylemma}
\begin{proof}
	The only changes made to the game graph were adding transitions to vertices $v \in V_1$ that belong to Spoiler.
	Since Spoiler always chooses a successor that increases the progress measure, if possible, the modification can only
	increase the progress measure of a vertex. This is because Spoiler has more possibilities to choose from
	after the modification.
	
	If the progress measure is unchanged the modification did not give Spoiler a better choice than before,
	if it increased it did create a better choice.
	
	Analogue for $\text{rem}(G^f_A, T)$ were Duplicator can only loose and
	not gain opportunities to decrease the measure. The functions \textit{add} and \textit{rem} make the
	game harder to win for Duplicator.\\
\end{proof}\quad\\
\lemmaref{lemma_pmincreases} allows us to initialize the data structures of every following run of \algoref{algo_jurdzinski}
with the values they had in the end of the first run. This can be done because we already know that Spoiler is capable of progressing
to this progress measures. It can only happen that he progresses further. By doing so we reduce the running
time by the part where the algorithm progresses to this state of progress measure.

In combination with the optimization \textbf{game graph history} we are even able to reuse the information gathered
from the last successful modification instead only from the initial run of \algoref{algo_jurdzinski}.

However, we need to take a closer look on the working list of the algorithm. Normally it gets initialized with vertices which have
a priority of $1$ but it can happen that we miss parts of the game graph. The modification, for example adding transitions,
may lead to the creation of new vertices $v \in V_0$ in the game graph. Imagine the addition of transition $(q, a, q')$,
this may create vertices $v_{(q, \tilde{q}, a)}$ if there was no from $q$ outgoing transition labeled with $a$ before.
If such a newly created vertex has a priority not equal to $0$ we also need to add it to the working list.

% Preprocessing
\subsubsection*{Preprocessing}
A requirement of \algoref{algo_jurdzinski} is that the input automaton neither has dead ends.
Since they do not contribute to the language of the automaton we safely remove them before
applying the algorithm by a linear search.

Furthermore this observation allows us to remove \textit{non live states} \libref{livestates} that do not contribute to the
language of a Büchi automaton too.
\begin{mydef}
	Given a Büchi automaton $A$ a state $q$ is \textnormal{live} iff
	\begin{align*}
		\exists w \in \mathcal{L}(A) \exists \text{ corresponding run } \pi : q \in \pi
	\end{align*}
	else $q$ is called a \textnormal{non live state}. I.e. $q$ occurs in an accepting run on some word.
\end{mydef}\quad\\
By doing so we reduce the size of the Büchi automaton and so the size of the game graph which often is a bottleneck.

Depending on the type of input automaton removing non live states may not always be a good idea, especially
when the costs for removing them is greater than the benefit of a smaller game graph.

% Fair-direct simulation
\subsubsection*{Fair-direct simulation}
Merging two states $q, q'$ is always possible without changing the language if they \textit{directly simulate}
each other (see \libref{simulation_general}). This also applies to \textit{delayed simulation}.

The idea for this optimization is to use delayed or direct simulation prior to fair simulation. Although delayed simulation,
in contrast to direct simulation, yields more merge-equivalent pairs, the costs for creating the huge game graph are high.
However, the game graph for direct simulation can be generated out of the graph for fair simulation
by omitting some edges, as seen in \defref{def_paritygamegraph}.

We generate the game graph for fair simulation $G^f_A$ and mark the edges needed to
omit for a direct game graph. Then we easily transform it to a direct game graph $G^{di}_A$,
the costs for the transformation are in $\mathcal{O}(1)$.
Next we apply direct simulation by using $\algoref{algo_jurdzinski}(G^{di}_A)$ and receive the relation $\preceq_{di}$.
After that we transform the graph back to $G^f_A$, again the costs are in $\mathcal{O}(1)$ only.
Now we start \algoref{algo_minimization}. As we reach \textbf{line 12}, where we check if
we can merge the states $q$ and $q'$ without changing the language, we first check if they directly simulate each other, i.e.
$q \preceq_{di} q' \land q' \preceq_{di} q$.
If that is the case, we already know that merging them does not change the language.
We skip applying an additional run of \algoref{algo_jurdzinski} and directly add the pair to $S_1$.

We can do likewise for transition removal. It holds that a transition of interest for removal $(q, a, q')$ can safely be
removed without changing the language if $\exists \tilde{q} \in Q : (q, a, \tilde{q}) \in \Delta \land q' \preceq_{di} \tilde{q}$,
compare to \sectionref{section_removetransitions}. I.e. if the required relation is direct simulation,
this was proven in \libref{fair_minimization}.

This optimization is more useful if more pairs of directly simulating vertices exist.

% Strongly connected components
\subsubsection*{Strongly connected components}
As we have seen \algoref{algo_jurdzinski} computes a bound, denoted by $\infty$, for the progress measure.
When a vertex reaches this bound one can be sure that there does not exist a winning strategy for Duplicator
when starting at this vertex.

Normally this bound is set to $n + 1$ where $n$ is the amount of vertices with a priority of $1$. When increasing to a
progress measure of $n + 1$ we have visited $n + 1$ vertices with priority $1$ without visiting priority $0$.
Then we are visiting at least one vertex with priority $1$ more than only once.

However, we can improve that bound for vertices locally. If, for example, we already know that a vertex can only reach
$5$ vertices with priority $1$ where there are $30$ in total, why not setting the local bound for this vertex to $6$ instead of $31$.
The computation for vertices would finish much faster in many applications.

This is where we use \textit{strongly connected components}, shortened to SCCs, lets take a look at the following definition.
\begin{mydef}
	A \textnormal{SCC} is a directed graph $G_S = \langle V, E\rangle$, where $V$ is the set of
	vertices and $E$ the set of edges, that is \textnormal{strongly connected}.
	$G_S$ is \textnormal{strongly connected} iff
	\begin{align*}
		\forall v \in V \forall v' \in V : v \text{ reachable from } v',
	\end{align*}
	i.e. every vertex in the SCC can be reached from every other vertex.
\end{mydef}\quad\\
We can use algorithms that work in linear time, for example Tarjan's algorithm \libref{scc}, to compute every SCC of
our game graph $G^f_A$ before we apply \algoref{algo_jurdzinski}. Now we compute for every vertex the local
optimal bound which is the number of vertices with priority $1$ that are in the SCC. Next we can use \algoref{algo_jurdzinski}
on each SCC separately.

However, the SCCs need to propagate progress measure updates among themselves.\\\\
As \algoref{algo_minimization} uses \algoref{algo_jurdzinski} multiple times while modifing the game graph and therefore
calculating the SCCs each time from scratch we may reuse information. Since the game graph gets only slightly changed,
we may already know how the SCCs change. If adding edges between SCCs they may merge to one SCC,
if removing edges in a SCC it may split into several SCCs.

We can either disable the SCC optimization on following runs of \algoref{algo_jurdzinski} and only use it on the first
run or we maintain SCCs of SCCs, i.e. calculating which SCCs can be reached from other SCCs.
If we have a SCC that contains several SCCs, we can merge them to one SCC.
As there are far less SCCs than vertices in the game graph, the SCC calculation of SCCs may be fast,
dependent on the type of input automaton.

% Order of vertex processing
\subsubsection*{Order of vertex processing}
\algoref{algo_jurdzinski} maintains the working list $L$, it contains vertices that need to be processed by the algorithm.
We can speed up the program by optimizing the order vertices get processed. The algorithm obviously
terminates the fastest if all vertices reach a progress measure of $\infty$ as fast as possible, if they can.
Therefore implementing the working list with a priority queue that first processes vertices with a higher
progress measure will significantly speed up the process.

If we detect that vertices frequently get added to the working list we should prioritize them over others
since chances are high they are part of a smaller loop which increases the progress measure faster
to $\infty$ than bigger loops.

% Game graph history
\subsubsection*{Game graph history}
When implementing \textbf{line 17, 23} of \algoref{algo_minimization} we should modify the original game
graph instead of creating a new game graph, since their size is big.
However, if an attempted merge or transition removal changes the language,
we must be able to revert changes to the game graph. Making a backup of the original graph may
not be a good idea, since they consume much space. But we can also memorize made changes and
implement an undo method.

Note that when using the optimization \textbf{reuse}, we must also be able to revert changes to the data
structures of \algoref{algo_jurdzinski} before reusing them in future runs of the algorithm.
Creating backups of the data structures needs space in the size of the game graph, $\mathcal{O}(|V|)$.
We should only memorize the exact changes that where made. For many vertices their value
in the data structures will remain unchanged since the game graph gets only slightly modified.

When a modifcation does not change the language we keep the modified graph and use this graph for next modifications.
This may slightly fasten the detection of future failing modifications, since the progress measure of vertices may
be greater than in the unmodified game graph and reach $\infty$ faster.

% Smart initialization
\subsubsection*{Smart initialization}
When applying \algoref{algo_jurdzinski} to a game graph we may already know that a state
$q'$ does not fairly simulate $q$. We then know for the vertex $v_{(q, q')}$ or $v_{(q, q', a)}$ that it will
reach $\infty$, we can speed the algorithm up by directly initializing $\mu(v_{(q, q')})$ with
$\infty$ and adding it to the working list to faster spread the progress measure in the game graph.

That is the case for vertices $v_{(q, q', a)} \in V_0$ if they are dead ends, they represent the position
before Duplicator's turn in the corresponding parity game.
Having no successor means loosing the game because Duplicator can not match the previous turn of Spoiler.
Note that there do not exist dead end vertices $v_{(q, q')} \in V_1$ because the algorithm
requires the input Büchi automaton to have no dead ends.

This optimization has great potential since many applications where the size of Büchi automata needs to be
reduced already have knownledge about simulation relation elements from the context.
For example the context may yield equivalence classes of states where simulation
between elements from different classes is impossible.

% Not reachable
\subsubsection*{Not reachable}
After removing edges in a game graph, there may exist vertices $v_{(q, q', a)} \in V_0$ that have no predecessors.
Those vertices are obsolete for the algorithm and can be removed to optimize future runs of the algorithm.

Note that vertices $v_{(q, q')} \in V_1$ should not get removed. Although they are not reachable from other
vertices, they represent a possible element of the simulation relation.

% Fast detection
\subsubsection*{Fast detection}
This optimization speeds up the compution if a modification changes the language,
i.e. when using \algoref{algo_jurdzinski} in \textbf{line 17, 23} of \algoref{algo_minimization}.

We can directly abort the computation of \algoref{algo_jurdzinski} as soon as one vertex
reaches a progress measure of $\infty$, while it has not reached it in the first run of \algoref{algo_jurdzinski}.
Then the simulation results of before and after the modification already differ.
We do not need to compute the rest of the simulation relation and directly know the modification changes the language.

% Equivalence classes
\subsubsection*{Equivalence classes}
In \textbf{lines 12-19} of \algoref{algo_minimization} when checking if a desired merge does change
the language we may often skip the process by using equivalence classes.

Take a look at the following example. We already know that the pair $q$ and $q'$ and the
pair $q'$ and $\tilde{q}$ can be merged without changing the language.
When now attempting the merge for $q$ and $\tilde{q}$ we can skip computing if it does change
the language because it will not. The state $\tilde{q}$ can be merged with $q'$, which can be merged with
$q$. Therefore $\tilde{q}$ and $q$ will also be mergeable without changing the language.

We detect such an equivalence class with a \textit{union-find data structure}. In this structure initially every
vertex has its own set. When a pair of states $q$ and $q'$ is mergeable without changing the language
we union their sets. Before attempting a merge we then check if both states are already in the
same set, if so we abort the process and already know they are safely mergeable.

% Experimental results
\section{Experimental results}
In this section we present experimental results for \algoref{algo_minimization}.
The results are compared with minimization techniques based on direct or delayed
simulation and two other minimization approaches.

Results where computed by a machine with an \textit{Intel Core i5-3570K} ($4\times3.40GHz)$ CPU.
The algorithms where written in \textit{Java}, the maximal heap size of the virtual machine
was restricted to \textit{10GB}.\\\\
We have implemented those methods in the \textit{AutomataLibrary} of the
ULTIMATE Project \libref{ultimate}, which is a program analysis framework.
Fair simulation minimization was implemented as described in \sectionref{section_minimization}
using some optimizations of \sectionref{section_optimization}. The implementation of direct and
delayed simulation minimization is similar to the approach shown in \libref{simulation_general}.
The other two techniques are called \textit{MinimizeSevpa} and \textit{ShrinkNwa}.
They are based on Hopcroft's algorithm and are described in \libref{minsevpashrinknwa} in more detail.
Further, we distinguish between fair simulation minimization and \textit{fair-direct} simulation
minimization. Fair-direct simulation is described in \sectionref{section_optimization}.

As before $A = \langle \Sigma, Q, Q_0, \Delta, F\rangle$ is a Büchi automaton and $G_A =\\
\langle V_0, V_1, E, p\rangle$ a game graph of $A$ where $V = V_0 \cup V_1$.
Further, $\infty = |\{v \in V : p(v) = 1\}| + 1$ as defined in \sectionref{section_computingterminology}.
The amount of states an algorithm has removed from the automaton is referred by $s$.
Analogously $t$ refers to the amount of removed transitions. The running time of an algorithm
is listed in the column \textit{time} and is measured in \textit{seconds}.

% Random automata
\subsection{Random automata}\label{section_randomautomata}
The automata set consists of $1000$ uniform distributed, random, connected automata.
The algorithm used for automata generation is desribed in \libref{randomdfa}.
They have $100$ states, an alphabet of size $5$ and $10$ final states.
% Tables
\begin{table}[h]
	 \begin{center}
		\begin{tabular}{|c|lrrrrrrr|}
			\hline
			\multicolumn{1}{|c|}{\cellcolor{black!30}\textbf{method}}
			&\multicolumn{1}{c}{\cellcolor{black!30}\textbf{time}}
			&\multicolumn{1}{c}{\cellcolor{black!30}$\bm{|Q|}$}
			&\multicolumn{1}{c}{\cellcolor{black!30}$\bm{|\Delta|}$}
			&\multicolumn{1}{c}{\cellcolor{black!30}$\bm{|V|}$}
			&\multicolumn{1}{c}{\cellcolor{black!30}$\bm{|E|}$}
			&\multicolumn{1}{c}{\cellcolor{black!30}$\bm{\infty}$}
			&\multicolumn{1}{c}{\cellcolor{black!30}\textbf{s}}
			&\multicolumn{1}{c|}{\cellcolor{black!30}\textbf{t}}\\
			\hline
			Fair			&$0.349$	&$100$	&$100$	&$22\,878$	&$11\,976$	&$901$	&$21$	&$0$\\
			Fair-Direct		&$0.736$	&$100$	&$100$	&$22\,878$	&$11\,976$	&$901$	&$21$	&$0$\\
			Delayed		&$0.09$	&$100$	&$100$	&$38\,800$	&$22\,963$	&$9\,001$	&$21$	&\emptyccellrightended\\
			Direct			&$0.065$	&$100$	&$100$	&$27\,724$	&$18\,020$	&$1$		&$0$	&\emptyccellrightended\\
			MinimizeSevpa	&$0.001$	&$100$	&$100$	&\emptyccell	&\emptyccell	&\emptyccell	&$20$	&\emptyccellrightended\\
			ShrinkNwa		&$0.001$	&$100$	&$100$	&\emptyccell	&\emptyccell	&\emptyccell	&$20$	&\emptyccellrightended\\
			\hline
		\end{tabular}
	\end{center}
	\caption{Experimental results averaged over $1000$ uniform distributed, random, connected automata.
		The alphabet size is $5$, the totality $5$\% and there are $10$ final states.}
	\label{random5tot}
	 \begin{center}
		\begin{tabular}{|c|lrrrrrrr|}
			\hline
			\multicolumn{1}{|c|}{\cellcolor{black!30}\textbf{method}}
			&\multicolumn{1}{c}{\cellcolor{black!30}\textbf{time}}
			&\multicolumn{1}{c}{\cellcolor{black!30}$\bm{|Q|}$}
			&\multicolumn{1}{c}{\cellcolor{black!30}$\bm{|\Delta|}$}
			&\multicolumn{1}{c}{\cellcolor{black!30}$\bm{|V|}$}
			&\multicolumn{1}{c}{\cellcolor{black!30}$\bm{|E|}$}
			&\multicolumn{1}{c}{\cellcolor{black!30}$\bm{\infty}$}
			&\multicolumn{1}{c}{\cellcolor{black!30}\textbf{s}}
			&\multicolumn{1}{c|}{\cellcolor{black!30}\textbf{t}}\\
			\hline
			Fair			&$0.125$	&$100$	&$250$	&$29\,701$	&$34\,801$	&$901$	&$1$	&$0$\\
			Fair-Direct		&$0.329$	&$100$	&$250$	&$29\,701$	&$34\,801$	&$901$	&$1$	&$0$\\
			Delayed		&$0.546$	&$100$	&$250$	&$57\,884$	&$67\,101$	&$9\,001$	&$1$	&\emptyccellrightended\\
			Direct			&$0.151$	&$100$	&$250$	&$44\,641$	&$45\,496$	&$1$		&$0$	&\emptyccellrightended\\
			MinimizeSevpa	&$0.001$	&$100$	&$250$	&\emptyccell	&\emptyccell	&\emptyccell	&$1$	&\emptyccellrightended\\
			ShrinkNwa		&$0.001$	&$100$	&$250$	&\emptyccell	&\emptyccell	&\emptyccell	&$1$	&\emptyccellrightended\\
			\hline
		\end{tabular}
	\end{center}
	\caption{Experimental results averaged over $1000$ uniform distributed, random, connected automata.
		The alphabet size is $5$, the totality $50$\% and there are $10$ final states.}
	\label{random50tot}
\end{table}\quad\\
In the first setup the totality of the automata is $5$\%, for the second it is $50$\%.
An automaton is \textit{total} iff every state has an outgoing transition for every letter of the alphabet,
i.e. $\forall q \in Q \forall a \in \Sigma \exists q' \in Q : (q, a, q') \in \Delta$.
\tableref{random5tot} and \tableref{random50tot} show the results.

By comparing the results, it strikes that fair simulation has the lowest running time for a totatility of $50$\%,
but turns out to be fairly bad for $5$\%. This is attributed to the size of the game graph which is far
bigger for direct or delayed simulation than for fair simulation. Additionally, there are less mergeable
states when increasing the totatility. Therefore, fair simulation minimization does apply \algoref{algo_jurdzinski} fewer
times for checking if a merge changes the language. As the game graph for fair simulation is small compared to the others,
a single run of \algoref{algo_jurdzinski} is very fast. The running time for fair simulation grows the more
often \algoref{algo_jurdzinski} needs to be applied.

Further, this setup shows that fair simulation in general does not remove more states than delayed
simulation for random automata. But fair simulation is a fast simulation based minimization solution for random
automata with high totatility.
% Table
\begin{table}[h]
	 \begin{center}
		\begin{tabular}{|c|lrrrrrrr|}
			\hline
			\multicolumn{1}{|c|}{\cellcolor{black!30}\textbf{method}}
			&\multicolumn{1}{c}{\cellcolor{black!30}\textbf{time}}
			&\multicolumn{1}{c}{\cellcolor{black!30}$\bm{|Q|}$}
			&\multicolumn{1}{c}{\cellcolor{black!30}$\bm{|\Delta|}$}
			&\multicolumn{1}{c}{\cellcolor{black!30}$\bm{|V|}$}
			&\multicolumn{1}{c}{\cellcolor{black!30}$\bm{|E|}$}
			&\multicolumn{1}{c}{\cellcolor{black!30}$\bm{\infty}$}
			&\multicolumn{1}{c}{\cellcolor{black!30}\textbf{s}}
			&\multicolumn{1}{c|}{\cellcolor{black!30}\textbf{t}}\\
			\hline
			Fair			&$0.15$	&$100$	&$300$	&$39\,107$	&$33\,064$	&$1\,601$	&$2$	&$0$\\
			Fair-Direct		&$0.365$	&$100$	&$300$	&$39\,107$	&$33\,064$	&$1\,601$	&$2$	&$0$\\
			Delayed		&$0.281$	&$100$	&$300$	&$74\,480$	&$60\,128$	&$8\,001$	&$2$	&\emptyccellrightended\\
			Direct			&$0.159$	&$100$	&$300$	&$65\,176$	&$50\,350$	&$1$		&$0$	&\emptyccellrightended\\
			MinimizeSevpa	&$0.001$	&$100$	&$300$	&\emptyccell	&\emptyccell	&\emptyccell	&$1$	&\emptyccellrightended\\
			ShrinkNwa		&$0.001$	&$100$	&$300$	&\emptyccell	&\emptyccell	&\emptyccell	&$1$	&\emptyccellrightended\\
			\hline
		\end{tabular}
	\end{center}
	\caption{Experimental results averaged over $1000$ uniform distributed, random, connected automata.
		The alphabet size is $30$, the totality $10$\% and there are $20$ final states.}
	\label{random30alphabet}
\end{table}\quad\\
In a third setup the size of the alphabet is $30$, the totatility $10$\% and the amount of final states is $20$.
\tableref{random30alphabet} shows the results.

Again, fair simulation is a good simulation based minimization strategy.
The reason is the same as before, there are not many states to remove and the game graph of fair simulation is much
smaller than for direct or delayed simulation.

When increasing the size of the alphabet for random automata there are less states to remove,
fair simulation benefits from this.

% Program analysis automata
\subsection{Program analysis automata}
The next set consists of automata obtained by applying termination
analysis with the \textit{Ultimate Büchi automizer} \libref{buchiautomizer} to \textit{C-programs}
from the benchmark set of the software verification competition \textit{SV-COMP 2016} \libref{svcomp16}.
In total the set has 485 automata.\\
% Table
\begin{table}[h]
	 \begin{center}
		\begin{tabular}{|c|lrrrrrrr|}
			\hline
			\multicolumn{1}{|c|}{\cellcolor{black!30}\textbf{method}}
			&\multicolumn{1}{c}{\cellcolor{black!30}\textbf{time}}
			&\multicolumn{1}{c}{\cellcolor{black!30}$\bm{|Q|}$}
			&\multicolumn{1}{c}{\cellcolor{black!30}$\bm{|\Delta|}$}
			&\multicolumn{1}{c}{\cellcolor{black!30}$\bm{|V|}$}
			&\multicolumn{1}{c}{\cellcolor{black!30}$\bm{|E|}$}
			&\multicolumn{1}{c}{\cellcolor{black!30}$\bm{\infty}$}
			&\multicolumn{1}{c}{\cellcolor{black!30}\textbf{s}}
			&\multicolumn{1}{c|}{\cellcolor{black!30}\textbf{t}}\\
			\hline
			Fair			&$1.503$	&$125$	&$189$	&$63\,086$	&$40\,461$	&$641$	&$35$	&$0$\\
			Fair-Direct		&$0.652$	&$125$	&$189$	&$63\,086$	&$40\,461$	&$641$	&$35$	&$0$\\
			Delayed		&$0.322$	&$125$	&$189$	&$102\,668$	&$43\,425$	&$1\,729$	&$33$	&\emptyccellrightended\\
			Direct			&$0.329$	&$125$	&$189$	&$101\,506$	&$77\,770$	&$1$		&$33$	&\emptyccellrightended\\
			MinimizeSevpa	&$0.002$	&$125$	&$189$	&\emptyccell	&\emptyccell	&\emptyccell	&$33$	&\emptyccellrightended\\
			ShrinkNwa		&$0.001$	&$125$	&$189$	&\emptyccell	&\emptyccell	&\emptyccell	&$33$	&\emptyccellrightended\\
			\hline
		\end{tabular}
	\end{center}
	\caption{Experimental results averaged over $114$ automata derived by the anaylsis of different programs.
		The automata have a size greater than $20$ and the set does not contain automata where no states can be removed.}
	\label{fullautomatanontrivial}
\end{table}\quad\\
The first setup is a subset which does not contain automata with a size smaller than $20$ and automata
where no states can be removed. The resulting set contains $114$ automata,
the outcome is shown in \tableref{fullautomatanontrivial}.

The fair simulation minimization algorithm is slow on this set.
The reason is the same as in \sectionref{section_randomautomata}. There are many states to remove,
thus fair simulation often applies \algoref{algo_jurdzinski} and the running time increases.

Though fair simulation is slow, the fair-direct optimization speeds up fair simulation significantly. This is because $33$ of
the removed $35$ states are direct simulation-equivalent states. As direct simulation minimization is computed in $0.329$ seconds,
the remaining fair simulation is computed in $0.652 - 0.329 = 0.323$ seconds when using the fair-direct optimization
instead of $1.503$ seconds.

The benefit of fair-direct simulation becomes more significant if there are more direct simulation-equivalent states. 
% Table
\begin{table}[h]
	 \begin{center}
		\begin{tabular}{|c|lrrrrrrr|}
			\hline
			\multicolumn{1}{|c|}{\cellcolor{black!30}\textbf{method}}
			&\multicolumn{1}{c}{\cellcolor{black!30}\textbf{time}}
			&\multicolumn{1}{c}{\cellcolor{black!30}$\bm{|Q|}$}
			&\multicolumn{1}{c}{\cellcolor{black!30}$\bm{|\Delta|}$}
			&\multicolumn{1}{c}{\cellcolor{black!30}$\bm{|V|}$}
			&\multicolumn{1}{c}{\cellcolor{black!30}$\bm{|E|}$}
			&\multicolumn{1}{c}{\cellcolor{black!30}$\bm{\infty}$}
			&\multicolumn{1}{c}{\cellcolor{black!30}\textbf{s}}
			&\multicolumn{1}{c|}{\cellcolor{black!30}\textbf{t}}\\
			\hline
			Fair			&$0.053$	&$72$	&$79$	&$19\,239$	&$11\,669$	&$858$	&$0$	&$0$\\
			Fair-Direct		&$0.142$	&$72$	&$79$	&$19\,239$	&$11\,669$	&$858$	&$0$	&$0$\\
			Delayed		&$0.805$	&$72$	&$79$	&$36\,951$	&$21\,515$	&$7\,944$	&$0$	&\emptyccellrightended\\
			Direct			&$0.077$	&$72$	&$79$	&$27\,790$	&$18\,400$	&$1$		&$0$	&\emptyccellrightended\\
			MinimizeSevpa	&$0.001$	&$72$	&$79$	&\emptyccell	&\emptyccell	&\emptyccell	&$0$	&\emptyccellrightended\\
			ShrinkNwa		&$0.001$	&$72$	&$79$	&\emptyccell	&\emptyccell	&\emptyccell	&$0$	&\emptyccellrightended\\
			\hline
		\end{tabular}
	\end{center}
	\caption{Experimental results averaged over $112$ automata derived by the anaylsis of different programs.
		The set does only contain automata where no states can be removed.}
	\label{fullautomatanoremove}
\end{table}\quad\\
The second subset only contains the automata where no states can be removed by neither of the
applied methods. These are $112$ automata in total, The results are shown in \tableref{fullautomatanoremove}.

Fair simulation has the smallest running time of all simulation based methods that where applied.
The reason why is the same as in \sectionref{section_randomautomata}.
The size of the fair simulation game graph and $\infty$ is much smaller than for delayed simulation.\\
% Table
\begin{table}[h]
	 \begin{center}
		\begin{tabular}{|c|lrrrrrrr|}
			\hline
			\multicolumn{1}{|c|}{\cellcolor{black!30}\textbf{method}}
			&\multicolumn{1}{c}{\cellcolor{black!30}\textbf{time}}
			&\multicolumn{1}{c}{\cellcolor{black!30}$\bm{|Q|}$}
			&\multicolumn{1}{c}{\cellcolor{black!30}$\bm{|\Delta|}$}
			&\multicolumn{1}{c}{\cellcolor{black!30}$\bm{|V|}$}
			&\multicolumn{1}{c}{\cellcolor{black!30}$\bm{|E|}$}
			&\multicolumn{1}{c}{\cellcolor{black!30}$\bm{\infty}$}
			&\multicolumn{1}{c}{\cellcolor{black!30}\textbf{s}}
			&\multicolumn{1}{c|}{\cellcolor{black!30}\textbf{t}}\\
			\hline
			Fair			&$0.223$	&$135$	&$196$	&$67\,195$	&$43\,089$	&$317$	&$1$	&$0$\\
			Fair-Direct		&$0.586$	&$135$	&$196$	&$67\,195$	&$43\,089$	&$317$	&$1$	&$0$\\
			Delayed		&$0.456$	&$135$	&$196$	&$108\,673$	&$46\,868$	&$2\,393$	&$0$	&\emptyccellrightended\\
			Direct			&$0.452$	&$135$	&$196$	&$107\,779$	&$83\,044$	&$1$		&$0$	&\emptyccellrightended\\
			MinimizeSevpa	&$0.003$	&$135$	&$196$	&\emptyccell	&\emptyccell	&\emptyccell	&$0$	&\emptyccellrightended\\
			ShrinkNwa		&$0.001$	&$135$	&$196$	&\emptyccell	&\emptyccell	&\emptyccell	&$0$	&\emptyccellrightended\\
			\hline
		\end{tabular}
	\end{center}
	\caption{Experimental results averaged over $485$ automata derived by the anaylsis of different programs.
		The method \textit{MinimizeSevpa} was applied on all automata prior to this experiment.}
	\label{fullautomataminsevpa}
\end{table}\quad\\
A similar result is yielded by the third subset. It contains all automata of the current set
after \textit{MinimizeSevpa} was applied on them. This setup is represented by \tableref{fullautomataminsevpa}.

As the size of the automata was already reduced by a minimization method, there are not many removable states anymore.
Fair simulation benefits from this and provides a fast simulation based technique for reducing the size of automaton even further.
For $12$ automata in this set, fair simulation even removed $40$ to $50$ states.

This setup shows that using a fast, coarser minimization technique before using fair simulation is a good strategy.
Moreover, it creates excellent prerequisites for the fair simulation algorithm.

% Complemented automata
\subsection{Complemented automata}\label{section_complementedautomata}
The automata sets in this subsection contain automata which are results of complementation algorithms.

Initially, we use the set of automata that was used in the following publication \libref{complementautomata}.
Then, for the results of \tableref{complementreducedoutdegree}, the algorithm \textit{ReducedOutdegree}
is used on the set. It is a \textit{rank-based} complementation algorithm which uses the
\textit{reduced average outdegree} optimization that is described in \textbf{Section~4} of \libref{reducedoutdegree}.

The results of \tableref{complementelastic} are derived analogously by using the algorithm \textit{Elastic}.
\textit{Elastic} is a yet unpublished Büchi complementation algorithm that is available in the
\textit{AutomataLibrary} of the ULTIMATE Project \libref{ultimate}.
% Tables
\begin{table}[h]
	 \begin{center}
		\begin{tabular}{|c|lrrrrrrr|}
			\hline
			\multicolumn{1}{|c|}{\cellcolor{black!30}\textbf{method}}
			&\multicolumn{1}{c}{\cellcolor{black!30}\textbf{time}}
			&\multicolumn{1}{c}{\cellcolor{black!30}$\bm{|Q|}$}
			&\multicolumn{1}{c}{\cellcolor{black!30}$\bm{|\Delta|}$}
			&\multicolumn{1}{c}{\cellcolor{black!30}$\bm{|V|}$}
			&\multicolumn{1}{c}{\cellcolor{black!30}$\bm{|E|}$}
			&\multicolumn{1}{c}{\cellcolor{black!30}$\bm{\infty}$}
			&\multicolumn{1}{c}{\cellcolor{black!30}\textbf{s}}
			&\multicolumn{1}{c|}{\cellcolor{black!30}\textbf{t}}\\
			\hline
			Fair			&$0.237$	&$19$	&$184$	&$1\,785$	&$6\,071$	&$120$	&$5$	&$31$\\
			Fair-Direct		&$0.345$	&$19$	&$184$	&$1\,785$	&$6\,071$	&$120$	&$5$	&$31$\\
			Delayed		&$0.042$	&$19$	&$184$	&$3\,056$	&$10\,192$	&$306$	&$4$	&\emptyccellrightended\\
			Direct			&$0.011$	&$19$	&$184$	&$4\,396$	&$8\,144$	&$1$		&$3$	&\emptyccellrightended\\
			MinimizeSevpa	&$0.001$	&$19$	&$184$	&\emptyccell	&\emptyccell	&\emptyccell	&$3$	&\emptyccellrightended\\
			ShrinkNwa		&$0.001$	&$19$	&$184$	&\emptyccell	&\emptyccell	&\emptyccell	&$3$	&\emptyccellrightended\\
			\hline
		\end{tabular}
	\end{center}
	\caption{Experimental results averaged over $104$ automata derived by the
		complementation algorithm \textit{ReducedOutdegree}.}
	\label{complementreducedoutdegree}
	 \begin{center}
		\begin{tabular}{|c|lrrrrrrr|}
			\hline
			\multicolumn{1}{|c|}{\cellcolor{black!30}\textbf{method}}
			&\multicolumn{1}{c}{\cellcolor{black!30}\textbf{time}}
			&\multicolumn{1}{c}{\cellcolor{black!30}$\bm{|Q|}$}
			&\multicolumn{1}{c}{\cellcolor{black!30}$\bm{|\Delta|}$}
			&\multicolumn{1}{c}{\cellcolor{black!30}$\bm{|V|}$}
			&\multicolumn{1}{c}{\cellcolor{black!30}$\bm{|E|}$}
			&\multicolumn{1}{c}{\cellcolor{black!30}$\bm{\infty}$}
			&\multicolumn{1}{c}{\cellcolor{black!30}\textbf{s}}
			&\multicolumn{1}{c|}{\cellcolor{black!30}\textbf{t}}\\
			\hline
			Fair			&$0.099$	&$16$	&$146$	&$1\,326$	&$4\,123$	&$93$		&$4$	&$17$\\
			Fair-Direct		&$0.159$	&$16$	&$146$	&$1\,326$	&$4\,123$	&$93$		&$4$	&$17$\\
			Delayed		&$0.032$	&$16$	&$146$	&$2\,434$	&$7\,104$	&$269$	&$3$	&\emptyccellrightended\\
			Direct			&$0.01$	&$16$	&$146$	&$3\,148$	&$5\,564$	&$1$		&$3$	&\emptyccellrightended\\
			MinimizeSevpa	&$0.003$	&$16$	&$146$	&\emptyccell	&\emptyccell	&\emptyccell	&$4$	&\emptyccellrightended\\
			ShrinkNwa		&$0.001$	&$16$	&$146$	&\emptyccell	&\emptyccell	&\emptyccell	&$4$	&\emptyccellrightended\\
			\hline
		\end{tabular}
	\end{center}
	\caption{Experimental results averaged over $102$ automata derived by the
		complementation algorithm \textit{Elastic}.}
	\label{complementelastic}
\end{table}\quad\\
Though fair simulation is $3$ to $5$ times slower than direct or delayed simulation, it removes many transitions of the automata.
This may even lead to states that become unreachable after the transition removal. Those states can also be removed,
therefore reducing the size of the automaton even further.

% Summary
\subsection{Summary}
The results show that for automata in general, one can not assume that minimization based on fair simulation
is faster than based on direct or delayed simulation.
However, if the automata have a small size or the expected amount of removed states is small, fair simulation
turns out to be the fastest, presented, simulation based method.

There are automata sets where fair simulation minimization can remove much more states than direct or delayed simulation minimization.
Additionally to the other methods, it also removes transitions.
For special automata sets, as seen in \sectionref{section_complementedautomata}, this is particularly effective.

Moreover, the space needed for fair simulation is much smaller compared to direct or delayed simulation.
This is attributed to the size of the game graph and can be seen in all experiments of this section.
Thus the presented algorithm can handle bigger automata where direct or delayed simulation methods run out of space.

% Conclusion
\section{Conclusion}
The presented algorithm reduces the size of Büchi automata based on fair simulation. We have seen that it is capable of merging
states and even removing redundant transitions. We also have seen that existing approaches using other simulation
relations are limited in their optimization capabilities, remove less redundancies and may even be more complex in time or space
for the case of delayed simulation.

A detailed description of the algorithm was presented as \algoref{algo_minimization} which allowed
us to develop optimization techniques as shown in \sectionref{section_optimization}.
Images and examples coming with the thesis illustrate sections which are more difficult to understand.
The thesis should made the algorithm comprehensible and the theory behind parity games and
the game graph be clear.\\\\
An aim for the future is to extend the algorithm for \textit{Nested Word automata}, presented in \libref{nwa}.
An equivalent, more easy to understand model are \textit{Visibly Pushdown automata}, as seen in \libref{vpa}.

This automata have a \textit{regular}, a \textit{call} and a \textit{return alphabet}.
Additionally they use a \textit{stack} to remember the \textit{call} levels.
If they, for example, use a transition \textit{call} $a$ they can use the transition \textit{return} $a$ after that.
If using transitions \textit{call} $a$, \textit{call} $a$, \textit{call} $b$ they may use \textit{return} $b$,
\textit{return} $a$, \textit{return} $a$ and so on. Transitions of the \textit{regular alphabet} can be used at anytime.
\textit{Nested Word automata} model this behavior without the use of a \textit{stack} by making the structure
of words, the \textit{nested words}, more complex.

The ability for an automaton to use different levels makes defining simulation relations and the construction
of a correct game graph tricky. However, also being able to reduce the size of \textit{Nested Word automata}
would be a great benefit for the ULTIMATE Project \libref{ultimate} and others as those automata describe
the behavior of programs closer than Büchi automata.

\clearpage

\bibliography{literature}
\end{document}